\documentclass[fleqn,usenatbib]{mnras}

\usepackage{newtxtext,newtxmath}
\usepackage[T1]{fontenc}

\DeclareRobustCommand{\VAN}[3]{#2}
\let\VANthebibliography\thebibliography
\def\thebibliography{\DeclareRobustCommand{\VAN}[3]{##3}\VANthebibliography}

\usepackage{graphicx}	% Including figure files
\usepackage{amsmath}	% Advanced maths commands
\usepackage{booktabs}   % Gives midruled lines in Table headers (nicer formatting)
\usepackage{xcolor}     % Allows colors for highlighting changes in response to reviewer comments

\title[\texttt{MANTA-Ray}: Supercharging Calculation Speeds in the Long-Wavelength Limit]{\texttt{MANTA-Ray}: Supercharging Speeds for Calculating the Optical Properties of Fractal Aggregates in the Long-Wavelength Limit}

\author[M. G. Lodge]{
M. G. Lodge$^{1}$\thanks{E-mail: m.g.lodge@bristol.ac.uk},
H. R. Wakeford$^{1}$,
Z. M. Leinhardt$^{1}$
\\
$^{1}$School of Physics, HH Wills Physics Laboratory, Tyndall Avenue, Bristol BS8 1TL, UK\\
}

\date{Accepted 2024 October 21. Received 2024 September 27; in original form 2024 August 01.}

\pubyear{2024}

\begin{document}
\label{firstpage}
\pagerange{\pageref{firstpage}--\pageref{lastpage}}
\maketitle

\begin{abstract}
Correctly modelling the absorptive properties of dust and haze particles is of great importance for determining the abundance of solid matter within protoplanetary disks and planetary atmospheres. Rigorous analyses such as the discrete dipole approximation (DDA) can be used to obtain accurate absorption cross-sections, but these require significant computing time and are often impractical to use in models. A simple analytical equation exists for spherical particles in the long-wavelength limit (where the wavelength is much larger than the size of the dust particle), but we demonstrate that this can significantly underestimate the absorption. This effect is found to depend strongly on refractive index, with values of $m=1+11$i corresponding to an underestimate in absorption by a factor of 1,000. Here we present \texttt{MANTA-Ray} (Modified Absorption of Non-spherical Tiny Aggregates in the RAYleigh regime): a simple model that can calculate absorption efficiencies within 10-20\% of the values predicted by DDA, but $10^{13}$ times faster. \texttt{MANTA-Ray} is very versatile and works for any wavelength and particle size in the long wavelength regime. It is also very flexible with regards to particle shape, and can correctly model structures ranging from long linear chains to tight compact clusters, composed of any material with refractive index 1+0.01i $\leq m \leq$ 11+11i. The packaged model is provided as publicly-available code for use by the astrophysical community.
\end{abstract}


%%%%%%%%%%%%%%%%%%%%%%%%%%%%%%%%%%%%%%%%%%%%%%%%%%

%%%%%%%%%%%%%%%%% BODY OF PAPER %%%%%%%%%%%%%%%%%%

\section{Introduction}

Optical models have become extremely important for determining structural and compositional information of a wide variety of astrophysical objects, for example moons and local solar system objects, protoplanetary discs, the interstellar medium, and exoplanet atmospheres. Knowing how particles can absorb and scatter light at certain wavelengths can reveal crucial diagnostic information about the atmosphere, temperature, formation mechanisms and a host of other important characteristics. In particular, calculated values for the masses of protoplanetary disks, or assumed optical extinction due to dust within a planetary atmosphere, depend strongly on the accuracy of the dust opacities that are used in models \citep[see reviews:][]{henning2013chemistry, gao2021aerosols}. However, for simplicity, particles are often assumed to be spherical, which can be a significant oversimplification.

\begin{figure}
    \includegraphics[width=\columnwidth]{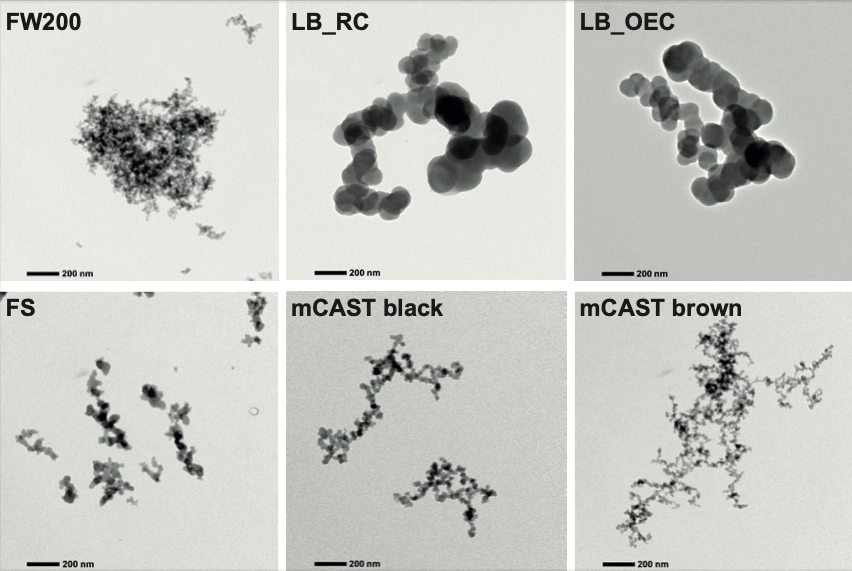}
    \caption{TEM images of atmospheric soot aggregates with a range of physicochemical properties and formation conditions. Reproduced with permission from \citet{mahrt2018ice}, Atmospheric Chemistry and Physics; published by EGU, 2018.}
    \label{fig:soot_TEM}
\end{figure}

\begin{figure}
    \includegraphics[width=\columnwidth]{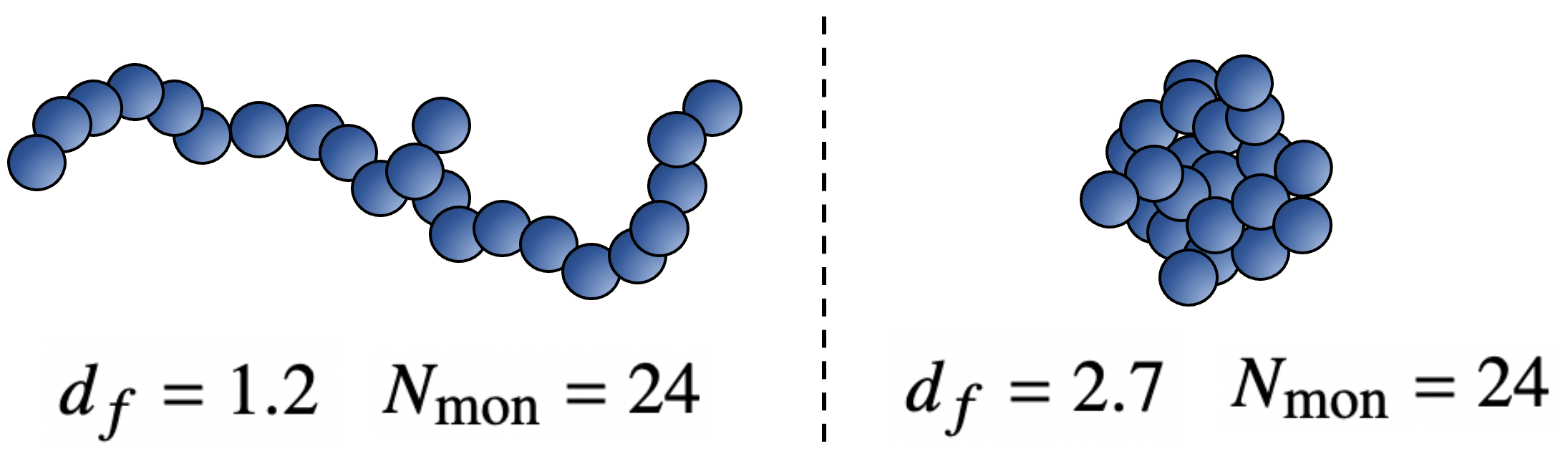}
    \caption{Two different aggregates composed of the same number of monomers, $N_\mathrm{mon}$, but with different fractal dimensions, $d_f$. In general, higher fractal dimensions lead to more compact shapes.}
    \label{fig:fractal_dimension}
\end{figure}

In reality, particles can form much more complex shapes, as seen in Transmission Electron Microscope (TEM) images of aerosols captured on Earth \citep{mahrt2018ice,katrinak1993fractal, yon2011examination, adachi2010shapes} (see Fig.~\ref{fig:soot_TEM}). Fractal aggregate aerosols often form clusters of roughly spherical monomers, which can be arranged in any shape from tight compact groups, to long linear branches, as a result of their formation conditions and chemical constituents. These fractal aggregates are often described by their fractal dimension $d_f$, which acts as a simple and convenient proxy to represent each shape type (Fig.~\ref{fig:fractal_dimension}). Although spherical particles are often used in models for simplicity, the optical properties of non-spherical particles (note that we use the term `particles' to refer to the entire aggregate, rather than just the monomers) have been shown to be very different to spheres in both experimental results \citep{munoz2004scattering, west1997laboratory} and a variety of different theoretical models \citep{mishchenko2009electromagnetic, bohren1991backscattering, draine1994discrete, hoshyaripour2019effects}. The shape of the dust particle has even been shown to have a significant impact on the amount of radiation absorbed within an atmosphere, and thus can effect energy balance of an entire planet \citep{wolf2010fractal}. For models that require mineral dust calculations in particular, \citet{nousiainen2009optical} strongly urges the community to avoid using Mie theory, because the use of spheres can be a major source of error in radiation-budget considerations. As such, there have been substantial efforts to improve the speed and accuracy of calculations for aerosols with more complex geometries than perfect spheres \citep[e.g.][]{kahnert2014model, yurkin2007discrete, mackowski1996calculation, lodge2024aerosols}.

The exact optical properties of particles depend on the wavelength, material type, particle structure and particle size. However, when the wavelength is much larger than the particle, calculations are significantly simplified, because the behaviour and electrostatic interactions between the radiation and material are much less complex in this regime. This long-wavelength limit can also be called the static-limit, or Rayleigh regime, in recognition for the initial studies performed by \citet{rayleigh1871light}. Its strict definition varies, but for the purposes of this study we define it as $\lambda \geq 100R$ (where $\lambda$ is the wavelength and $R$ is the radius of a spherical particle of equivalent mass). 

\citet{bohren2008absorption} describe how for spheres in this regime, the series of complex Mie formula that are usually used to calculate absorption can be reduced to a single analytical equation. A simple analytical analogue for non-spherical particles has not been found, owing to the complexities in solving analytical equations for more complex geometries. This means that calculating the optical properties of a single non-spherical particle can involve solving groups of over a million linear equations, which dramatically increases computation time and makes their utilisation in larger models challenging (see section \ref{section:DDA}). However, progress has been made in accelerating calculations; \citet{min2006absorption} discovered a method that significantly decreases the computation time in the long-wavelength limit, by exploiting matrix symmetries that appear in the discrete dipole approximation for small size parameters. This means that a rigorous calculation is only required once for each particle shape, and one can then calculate the optical properties for any refractive index, wavelength and orientation, dramatically decreasing the time required to study a wide range of wavelengths or particle sizes. In addition, \citet{farias1996range} showed that for optically soft particles (with small refractive index $m$, such that $|m-1|<1$), a version of Rayleigh-Debye-Gans developed for fractal aggregates (RGD-FA) has been found to achieve results with errors of less than 30\%.

In this paper, we explore a particular phenomenon that occurs specifically in this limit; a significant enhancement in the amount of absorption that fractal aggregates exhibit in comparison to spheres of the same mass/volume. This enhancement has been noticed previously, and is the proposed explanation for large increases in absorption for interstellar dust grains \citep{wright1987long, bazell1990effects}. \cite{henning1995dust} provides an intuitive qualitative justification for the effect by approximating long chains of spheres as cylinders. \citet{fogel1998modeling, kohler2011aggregate} used the discrete dipole approximation \citep{purcell1973scattering,draine1994discrete} to demonstrate the increased absorption in fractal aggregates. In addition, \citet{stognienko1995optical} compared several optical models (effective medium theory \citep{maxwell1904xii}, the discrete dipole approximation and the discrete multipole method \citep{sheu1990simulation, hinsen1992dielectric} for aggregates formed by particle-cluster and cluster-cluster agglomeration, and found higher absorption in aggregates (by factors of between 2 and 1,000) when compared to spheres of the same mass. \citet{liu2013investigation} studied soot aggregates with a variety of fractal dimensions, using the Rayleigh-Debye-Gans (RGD) approximation and the Generalised Multiparticle Mie solution (GMM) \citep{xu1995electromagnetic}, and found large enhancements for particles that were small compared to the wavelength. Other authors have found similar results, using a variety of other optical models \citep{mackowski1995electrostatics, mackowski2006simplified}. Here, we attempt to create a fast, general and predictive analytical equation that estimates this enhancement, in significantly less time than all models above. We present a non-spherical analogue of the equation for spherical particles -- a simple model that can predict optical properties of fractal aggregates for almost any material composition, particle size and wavelength within the long-wavelength limit.

In section \ref{section:theory}, we outline the theory used to determine the optical properties of both spherical and non-spherical particles in this paper, as well as outlining our novel simplified approximation model. In section \ref{section:method}, we describe our methodology for generating realistic geometries of non-spherical particles. In section \ref{section:results}, we use a set of 22 dust particle shapes to define the parameters of our model, and then another set of 18 particle shapes to test it's performance. In section \ref{section:conclusions} we summarise the average and maximum errors expected from use of the model, detail its range of applicability, and suggest use cases for the astrophysical community.

\section{Theory} \label{section:theory}

When analysing aerosol particles, we typically aim to determine three optical properties: the absorption efficiency $Q_\mathrm{abs}$, the scattering efficiency $Q_\mathrm{sca}$, and the extinction efficiency $Q_\mathrm{ext}$. They are linked by:
\begin{equation}
    Q_\mathrm{ext} = Q_\mathrm{abs} + Q_\mathrm{sca}.
	\label{eq:Q_ext}
\end{equation}
However, in the long--wavelength limit, $Q_\mathrm{abs} >> Q_\mathrm{sca}$ and so we can use the approximation:
\begin{equation}
    Q_\mathrm{ext} \approx Q_\mathrm{abs}.
\end{equation}
The absorption and extinction efficiencies in this regime are thus interchangeable, but we will consistently refer to absorption in this paper, as a reminder of the specific mechanism of extinction. Once the absorption efficiency is obtained for a certain particle, the cross-section can be found from:
\begin{equation}
    C_\mathrm{abs} = Q_\mathrm{abs} \pi R^2,
	\label{eq:cross_section}
\end{equation}
where $R$ is the mass-equivalent radius of the particle (the radius of a sphere of the same mass). 

\subsection{Spherical particles}
For spherical particles, the absorption efficiency can be determined using Mie theory \citep{mie1908beitrage,bohren2008absorption}, which gives exact analytical solutions for Maxwell's equations applied to a perfect homogeneous sphere. \citet[see their sec. 5.2]{bohren2008absorption} demonstrate that by expanding the power series expansions of spherical Bessel functions in the long-wavelength limit, a number of very complex Mie formulae can be reduced to a simple, single equation:
\begin{equation}
    Q_\mathrm{abs,sphere} = \frac{8\pi R}{\lambda}\mathrm{Im} \left( \frac{m^2 -1}{m^2+2} \right),
	\label{eq:Rayleigh_original}
\end{equation}
where $R$ is the radius of the particle, $\lambda$ is the wavelength of radiation and $m$ is the refractive index, which includes a real ($n$) and imaginary ($k$) component such that $m=n+k\mathrm{i}$. This simple formula is often used because it is much quicker than calculating the coefficients and summations for Mie theory (see Eq. 1-5 of \citet{lodge2024aerosols} for comparison), but it agrees to at least 4 significant figures for all values studied in this paper ($m \leq 11 + 11\mathrm{i}$).

\subsection{Non-spherical particles}

\subsubsection{General concept of the model: \texttt{MANTA-Ray}}

\begin{figure}
    \includegraphics[width=\columnwidth]{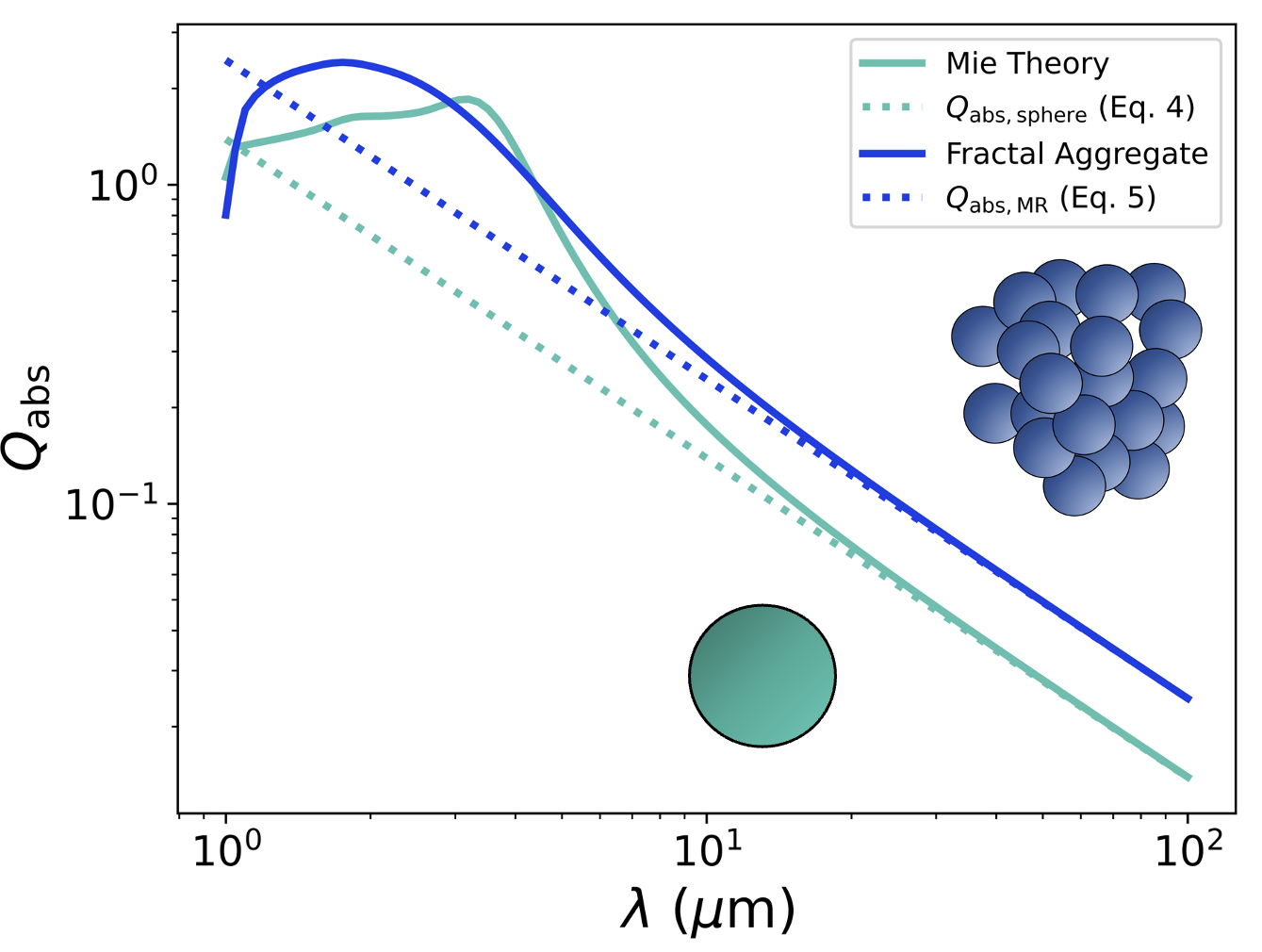}
    \caption{Predicted values of absorption efficiency $Q_\mathrm{abs}$ for a realistic aerosol shape (solid blue line, labelled \lq{Fractal Aggregate}\rq, calculated using DDA), versus a sphere of the same mass (solid teal line, \lq{Mie theory}\rq). Also shown is an approximation for Mie theory in the long wavelength limit (dashed teal line, $Q_\mathrm{abs, sphere}$), and the \texttt{MANTA-Ray} approximation for fractal aggregates presented in this paper (dashed blue line, $Q_\mathrm{abs, MR}$). All models use a volume-equivalent particle radius of $R=0.5~\mu$m and a refractive index of $m=3+0.8i$. The aggregate shown has fractal dimension $d_f=2.5$, and represents a realistic compact cluster shape (composed from 3D-TEM images of an Earth-based soot aerosol captured above Mexico city by \citet{adachi2010shapes}).}
    \label{fig:DDA_vs_Mie_vs_Rayleigh}
\end{figure}

In the long-wavelength limit, Mie theory can significantly underestimate the amount of absorption in non-spherical particles (see Fig.~\ref{fig:DDA_vs_Mie_vs_Rayleigh}). In this paper, we demonstrate that we can predict the true absorption efficiencies for almost any complex-shaped fractal aggregate by adding a simple multiplicative factor to Eq.~\ref{eq:Rayleigh_original}:

\begin{equation}
    Q_\mathrm{abs, MR} = \chi(n,k,d_f)\frac{8\pi R}{\lambda}\mathrm{Im} \left( \frac{m^2 -1}{m^2+2} \right).
	\label{eq:Modified_Rayleigh}
\end{equation}

We have determined that the multiplicative factor $\chi$ is a function of the real and imaginary components of refractive index ($n$ and $k$ respectively), as well as the fractal dimension of the particle $d_f$. We introduce this model of absorption as \texttt{MANTA-Ray} (Modified Absorption for Non-spherical Tiny Aggregates in the Rayleigh regime)\footnote{\url{https://github.com/mglodge/MANTA-Ray}}.

The effect of the multiplicative term is demonstrated in Fig.~\ref{fig:DDA_vs_Mie_vs_Rayleigh}. For a perfectly spherical $R=0.5~\mu$m particle, Eq.~\ref{eq:Rayleigh_original} (dotted green line) converges to agree with the correct solutions for Mie theory (solid green line) above wavelengths of $\approx20~\mu$m. As expected, Eq.~\ref{eq:Rayleigh_original} is only applicable in this long-wavelength limit, because the simple $Q_\mathrm{abs}~\propto~\lambda^{-1}$ relationship cannot represent the complex interactions that occur at small wavelengths (when $\lambda$ is close to the particle size). Conversely, for the fractal aggregate shown, the true absorption is calculated (using DDA -- see section \ref{section:DDA}) to be much higher than for spheres of the same mass. Interestingly, the true absorption is also a function of $\lambda^{-1}$ in the long-wavelength limit. Therefore, $\chi$ in Eq.~\ref{eq:Modified_Rayleigh} acts as an offset that multiplies Eq.~\ref{eq:Rayleigh_original} to give the correct absorption for non-spherical shapes, and it does so tens of orders of magnitude faster than any other methods of calculating the increase in absorption for non-spherical particles rigorously (see section \ref{section:DDA}). The offset shown in this case (Fig.~\ref{fig:DDA_vs_Mie_vs_Rayleigh}) is a factor of $\approx2$, but in section \ref{section:results} we demonstrate that it can be factors of 1,000 for higher refractive indices and shapes that are more elongated.

\subsubsection{Rigorously calculating the optical properties of non-spherical particles: DDA} \label{section:DDA}

There are a handful of solutions for determining the optical properties of simple non-spherical shapes (e.g. infinite cylinders, hollow spheres etc). However, for a general shape that could have any potential geometry, there are two benchmark models that are widely used to give exact solutions -- DDA (the discrete dipole approximation) \citep{purcell1973scattering, draine1994discrete} and T-matrix \citep{waterman1965matrix, mishchenko1996t}. The former was chosen for this research because of it's flexibility in allowing the option of heterogeneous materials, and because it has been well-tested for the extreme refractive indices explored in this paper.

DDA essentially breaks a particle into $N$ discretised cubes, and determines the scattering effects of electric fields interacting with each possible pair of cubes. The name can be considered misleading, because it is really an exact theory in the limit $N\rightarrow \infty$. However, in practice the accuracy of DDA is limited by computing resources. Computation time for a particular calculation is highly dependent on $N$, due to the necessity to solve increasing numbers of linear equations to achieve higher accuracies (see \citealt{lodge2024aerosols, draine1994discrete} for a detailed description of the problem and also example calculation times for realistic aerosols of various discretisation values of $N$). We follow the same general method in this paper, but use a different prescription of DDA. Crucially, to enable exploration of higher values of refractive indices (up to $m = 11 + 11\mathrm{i}$), we use the Filtered-Coupled Dipole method \citep{yurkin2010application, piller1998increasing, gay2002library} instead of the LDR prescription. The refractive index range was chosen to encompass a wide range of extreme refractive indices that materials can exhibit in the infra-red, for example the condensates studied in \citet{wakeford2015transmission} (see Fig.~1 in particular). In addition, FCD has been historically well-tested within these limits.

To summarise the computations that are required for DDA, for a particle composed of $N$ dipoles, we aim to solve the following series of $3N$ linear equations:
\begin{equation} \label{eq:compact_linear}
    \sum_{k=1}^{N}\mathbf{G_{jk}P_{k}}=\mathbf{E_{inc,j}},
\end{equation}
where $\mathbf{E_{inc,j}}$ represents the incident electric field at dipole $j$ (with the same conventions for rotations and polarisation state as in Eq. 30 and 31 of \citealt{lodge2024aerosols}), and $\mathbf{P_{k}}$ is the polarisation of each of the other dipoles $\mathrm{\mathbf{k}}$. We then define $\mathbf{G_{jk}}$ as a tensor that calculates the contribution of the scattered electric field from all other dipoles $\mathrm{\mathbf{k}}$ at dipole $\mathrm{\mathbf{j}}$. For diagonal elements where $\mathrm{\mathbf{j}}=\mathrm{\mathbf{k}}$, we define:
\begin{equation} \label{eq:G_jj}
    \mathbf{G_{jj}} = \frac{1+D}{\alpha_\mathrm{CM}}.
\end{equation}
where $D$ is given by:
\begin{equation} \label{eq:alpha_FCD}
    D= \frac{\alpha_\mathrm{CM}}{d^3} \left( \frac{4}{3}(kd^2) + \frac{2}{3\pi} \ln \left[ \frac{\pi - kd}{\pi + kd} \right] (kd^3) - \frac{2}{3}i(kd)^3 \right)
\end{equation}
and $d$ is the spacing between dipoles, and $k=\frac{2\pi}{\lambda}$ (to keep consistent with other literature we have used notation $k$ for wavenumber, and for the imaginary component of refractive index, as well as $\mathbf{k}$ for dipole index, but we highlight that these are not the same and care should be taken. Discussions of wavenumber and dipole index are restricted to this theory section only -- all other instances of $k$ refer to the imaginary component of refractive index). The Clausius-Mossotti relation $\alpha_\mathrm{cm}$ expresses the dielectric constant in terms of the polarisability of a material's constituent dipoles:
\begin{equation}
    \alpha_\mathrm{cm}=\frac{3d^{3}}{4\pi}\frac{\epsilon-1}{\epsilon+2},
\end{equation}
where $\epsilon$ is the dielectric function of the aerosol (and $\epsilon=m^{2}/\mu$, with $\mu$ as the relative permeability of the material).

The non-diagonal elements of the matrix $\mathbf{G_{jk}}$ are given by:
\begin{equation} \label{G_jk_equation}
    \mathbf{G_{jk}} = \begin{pmatrix} G_{xx} & G_{xy} & G_{xz} \\ G_{yx} & G_{yy} & G_{yz} \\ G_{zx} & G_{zy} & G_{zz} \end{pmatrix},
\end{equation}
where
\begin{equation}
    G_{uu} = (kd)^2g_0 + \frac{g_1}{r} + \frac{2}{3}h^r + \left(\frac{g_2}{r^2} - \frac{g_1}{r^3} \right)uu,
\end{equation}
and
\begin{equation}
    G_{uv} = \left(\frac{g_2}{r^2} - \frac{g_1}{r^3} \right)uv.
\end{equation}
Terms $x$, $y$ and $z$ are the relative positions of the dipoles in Cartesian coordinates, and $r=\sqrt{x^{2}+y^{2}+z^{2}}$. The other terms above are given by:
\begin{equation}
    g_{0} = \frac{ \left( \exp(ikR)-\frac{\alpha_0}{\pi} \right) }{R},
\end{equation}
\begin{equation}
    g_{1} = \frac{ \left( \exp(ikR) \left[ ikR-1] \right] -\frac{\alpha_0 - \alpha_1 R}{\pi} \right) }{R^2},
\end{equation}
\begin{equation}
    g_{2} = \frac{ \left( \exp(ikR) \left[ 2 - 2ikR - (kR)^2] \right] -\frac{2 (\alpha_0 - \alpha_1 R) + R^2 \alpha_2}{\pi} \right) }{R^3},
\end{equation}
\begin{equation}
    \alpha_{0} = \sin(kR) [ \mathrm{Ci}^+ - \mathrm{Ci}^- ] + \cos(kR) [ \pi - Si^+ - Si^- ],
\end{equation}
\begin{equation}
    \alpha_{1} = k\sin(kR) [ -\pi + \mathrm{Si}^+ - \mathrm{Si}^- ] + k\cos(kR) [ \mathrm{Ci}^+ - \mathrm{Ci}^- ] - \frac{2 \sin(k_{F}R)}{R},
\end{equation}
\begin{multline}
    \alpha_{2} = k^2\sin(kR)[ \mathrm{Ci}^- - \mathrm{Ci}^+ ] + k^2\cos(kR) [ \mathrm{Si}^- + \mathrm{Si}^+ - \pi ] - \frac{2 \sin(k_{F}R)}{R^2}  \\ - \frac{2 k_{F}\cos(k_{F}R)}{R},
\end{multline}
\begin{equation}
    h^{r} = k\sin(kR) [ -\pi + \mathrm{Si}^+ - \mathrm{Si}^- ] + k\cos(kR) [ \mathrm{Ci}^+ - \mathrm{Ci}^- ] - \frac{2 \sin(k_{F}R)}{R}.
\end{equation}
The optimal value for parameter $k_F = \pi$ (see \citet{gay2002library} for more details). The terms $\mathrm{Si}^\pm =\mathrm{Si}[(\pi \pm k)R]$ and $\mathrm{Ci}^\pm =\mathrm{Ci}[(\pi \pm k)R]$ represent the Sine and Cosine integrals respectively, which can be evaluated efficiently using the Pad\'e approximants and Chebyshev-Pad\'e expansions given in \citet[][see their Appendix B]{rowe2015galsim}.

Once the series of linear equations has been solved and the values of $\mathbf{P}$ in Eq. \ref{eq:compact_linear} have been found, we calculate extinction efficiency using:
\begin{equation} \label{DDA_Q_abs}
Q_\mathrm{abs, DDA}=\frac{4k}{R^2 \vert E_{0} \rvert^{2}} \sum_{j=1}^{N} \{\mathrm{Im}(P_{j}.(\alpha_{j}^{-1})^{*}P_{j}^{*}) - \frac{2}{3}k^{3}\lvert P_{j} \rvert^{2}\}.
\end{equation}
In practice the electric field is normalised such that $\vert E_{0} \rvert^{2}=1$ (the initial polarisation state of the incoming radiation is chosen as a unit vector), and $R$ is the radius of a sphere which would have the same volume as the non-spherical shape.

\subsubsection{Validity}

For DDA to be considered accurate, the two following criteria are often quoted as being required to be satisfied \citep{draine1994discrete}:

\begin{enumerate}
    \item the dipole size is small compared to the wavelength: $|m|kd<1$
    \item $d$ must be small enough to describe the geometry satisfactorily.
\end{enumerate}

Because this paper studies the long-wavelength limit, condition (i) is easily met for all aggregates studied in this paper. Condition (ii) is harder to define mathematically, however it can be checked by further discretising particles to see whether the results of DDA have converged \citep{yurkin2010application, yurkin2006convergence, liu2018performance}. To do this, we increased the resolution of shapes with fractal dimensions of 1.2 and 2.7 until they no longer varied by more than a few percent for a variety of refractive indices within the regime studied, including for the extremes (11 + 11i). This convergence occurred by $N\approx65,000$, and so all shape files in this study have at least this number of dipoles.

In addition to the above criteria, values obtained using DDA are often determined at several different dipole resolutions, and then extrapolated to $N \rightarrow \infty$ to obtain highly accurate values \citep{shen2008modeling, draine2016graphite}. However, given that these extrapolation errors are typically small for such high dipole numbers, and due the large number of different refractive indices and shapes studied here, we have opted not to perform extrapolation of these results in this case due to computational limitations.

\section{Method} \label{section:method}

\subsection{Non-spherical particle behaviour in long-wavelength limit}

\begin{table}
    \centering
	\caption{Absorption efficiencies calculated for a sphere ($Q_\mathrm{abs,sphere}$) of radius $R=0.5~\mu m$ and $m=11+11\mathrm{i}$ at a variety of wavelengths. We compare this to absorption efficiencies calculated for two different fractal aggregates of identical mass to the sphere, but different shapes (with fractal dimensions of 1.2 and 2.7, respectively) using DDA ($Q_\mathrm{abs,DDA}$). Values for $\chi$ are calculated using Eq.~\ref{eq:chi}.}
	\label{table:proof_of_concept}
    \setlength{\tabcolsep}{4pt} % reduce white space to fit table onto one page
    \begin{tabular}{r c c c c c} % NOTE: if we ever need more than 10 columns, just add the number in curly braces! e.g. the {12} here.
    \hline
    \multicolumn{2}{c}{} &  \multicolumn{2}{c}{Shape 1.2a} & \multicolumn{2}{c}{Shape 2.7a} \\
    \cmidrule(lr){3-4} \cmidrule(lr){5-6}
    $\lambda (\mu m)$   & $Q_\mathrm{abs,sphere}$ & $Q_\mathrm{abs,DDA}$ & $\chi$ & $Q_\mathrm{abs,DDA}$ & $\chi$ \\
    \hline
        100 & $1.558\times 10^{-3}$ & $5.092\times 10^{-1}$ & 326.9 & $1.028\times 10^{-1}$ & 65.0 \\
      1,000 & $1.568\times 10^{-4}$ & $5.081\times 10^{-2}$ & 324.1 & $1.020\times 10^{-2}$ & 65.0 \\
     10,000 & $1.558\times 10^{-5}$ & $5.081\times 10^{-3}$ & 326.1 & $1.020\times 10^{-3}$ & 65.4 \\
    100,000 & $1.558\times 10^{-6}$ & $5.081\times 10^{-4}$ & 326.1 & $1.020\times 10^{-4}$ & 65.4 \\
    \hline
    \end{tabular}
\end{table}

From Eq.~\ref{eq:Rayleigh_original}, it can be seen that $Q_\mathrm{abs} \propto \frac{1}{\lambda}$ (for spherical particles), and this result is well-known. However, one of the interesting observations of this paper is that this relationship is also true for fractal aggregates in the long-wavelength limit. This point is demonstrated visually in Fig.~\ref{fig:DDA_vs_Mie_vs_Rayleigh}, and we demonstrate this numerically in Table~\ref{table:proof_of_concept} for a wider range of wavelengths, for two shapes of very different fractal dimension values: 1.2 (long and linear) and 2.7 (a compact cluster). We suggest that for aggregates of any shape, the modification term $\chi$ depends strongly on fractal dimension, but appears to be invariant as a function of wavelength (or equivalently, independent of size parameter, given by $\frac{2 \pi R}{\lambda}$). To study the form of $\chi$ in detail, we calculate it for a range of 22 different fractal aggregates, create a model of it's variation as a function of fractal dimension, and then test this model on a further 18 aggregates of different particle size and wavelength. In each case we calculate $\chi$ as the ratio of the absorption calculated by DDA for the real shape and the absorption predicted for a spherical particle of equivalent volume:
\begin{equation} \label{eq:chi}
    \chi(n,k,d_f) = \frac{Q_\mathrm{abs, DDA}}{Q_\mathrm{abs, sphere}}.
\end{equation}
Dipole spacing $d$ can be calculated for each shape (with each described by a different number of dipoles $N$) using conservation of volume between the spherical and discretised models:
\begin{equation}
    \frac{4}{3}\pi R^3 = N d^3.
\end{equation}

\subsection{Particle shapes} \label{section:particle_shapes}

\begin{figure*}
    \includegraphics[width=\textwidth]{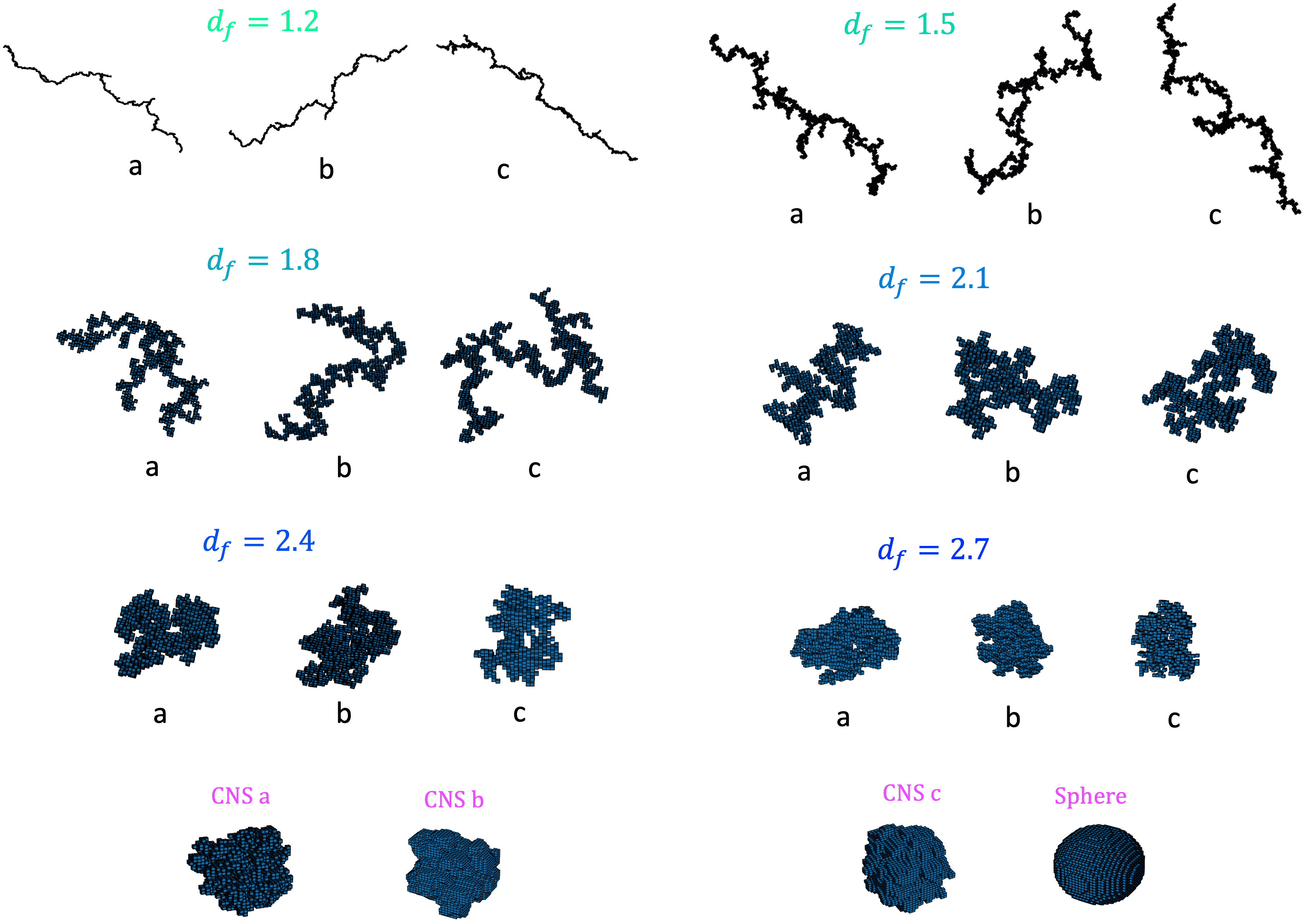}
    \caption{Aggregates used to create the model. CNS aggregates are the most compact, and clusters with fractal dimensions of 1.2 are the least compact.}
    \label{fig:training_fractals}
\end{figure*}

Realistic particle shapes were generated using \texttt{aggregate gen} \citep{aggregate_gen}, an efficient C++ code for creating cluster-cluster aggregates (CCA) of spheres, based on the methods of \citet{filippov2000fractal}. We created an initial set of 18 clusters (Fig.~\ref{fig:training_fractals}), each composed of 1,024 spherical monomers, with a range of fractal dimensions between 1.2 (very long and elongated chains) to 2.7 (almost spherical particles, but still quite porous). The fractal dimension represents the `compactness' of the cluster (Fig.~\ref{fig:fractal_dimension}), linking the number of monomers (spheres that make up the cluster) $N_\mathrm{mon}$ and the radius of the monomers $R_\mathrm{mon}$ to the radius of gyration $R_\mathrm{g}$ (the radial distance to a point that would have the same moment of inertia as the cluster, if a point mass with the same mass as the cluster were placed there):
\begin{equation} \label{fractal_equation}
    N_\mathrm{mon}=k_{0} \left(\frac{R_{g}}{{R_0}}\right)^{d_{f}}.
\end{equation}
$k_0$ is the fractal prefactor (a scaling relation), which we determine using the approximation given in \citet{Tazaki_2021} (Eq. 2, which is correct to within 5\% for all of the structure-types explored here):
\begin{equation} 
    k_{0} \approx 0.716(1-d_{f}) + \sqrt{3}.
\end{equation}

In addition, we studied three much more compact irregular shapes, using publicly-available\footnote{\url{https://www.astro.princeton.edu/~draine/agglom.html}} files created by the authors of \citet{shen2008modeling}, using the ballistic and migration (BAM2) methodology outlined within. In summary, BAM2 aggregates are created by particles that move at random trajectories, collide with a cluster, and then can roll (or \lq{migrate}\rq) to a new position, which creates very compact aggregates. The aggregates studied in this paper are then even more compacted, because of the nature of discretising the positions into a cubical grid. They therefore act as compact non-spherical (CNS) particle examples for this study -- they have a lower porosity than regular BAM aggregates, and very few spaces in between the dipoles.

Finally, we also created a pseudosphere (a discretised sphere, made of small cubical dipoles), which we expect to show very little deviation from the spherical model when DDA is applied to it. That is, we know that we should expect $\chi \approx 1$, acting as a helpful \lq{sanity check}\rq~that the results of DDA are valid for the particle radii, wavelengths, and values of $m$ explored.

All methods above initially generated coordinates of roughly 1,000 dipoles, and so to increase the resolution and retain realistic spherical edges, we used two iterations of \texttt{SPHERIFY} (a purpose built code that upscales the resolution of fractal aggregates -- see \citet{lodge2024aerosols} for details and methodology). This process resulted in each aggregate being represented by $\approx70,000$ dipoles in total, roughly 70 dipoles per monomer. Our dipole positions are based on the assumption that monomers are spherical, but even if they were slightly elliptical, the discretisation of the grid at this resolution would place dipoles at almost identical positions, so the model can be assumed to be valid for these cases. However, it may not sufficiently represent aggregates composed of smaller monomers that are more exotic shapes, for example, extremely elongated ellipsoids.

We checked for convergence of the number of angles required for a consistent orientational average in the same way as in \citet{lodge2024aerosols} -- by calculating the orientational average using increasing numbers of angles from a recurring icosahedral grid, including for the most extreme shapes ($d_f=1.2$ and 2.7) and all refractive indices studied. We found that an orientational and polarisation average that using 12 angles and 2 polarisation states was enough to obtain converged results to within a few percent. Combined with the discretisation convergence tests, we therefore assume that DDA is valid for the particles and regime studied here.

\section{Results} \label{section:results}

\begin{figure}
    \includegraphics[width=\columnwidth]{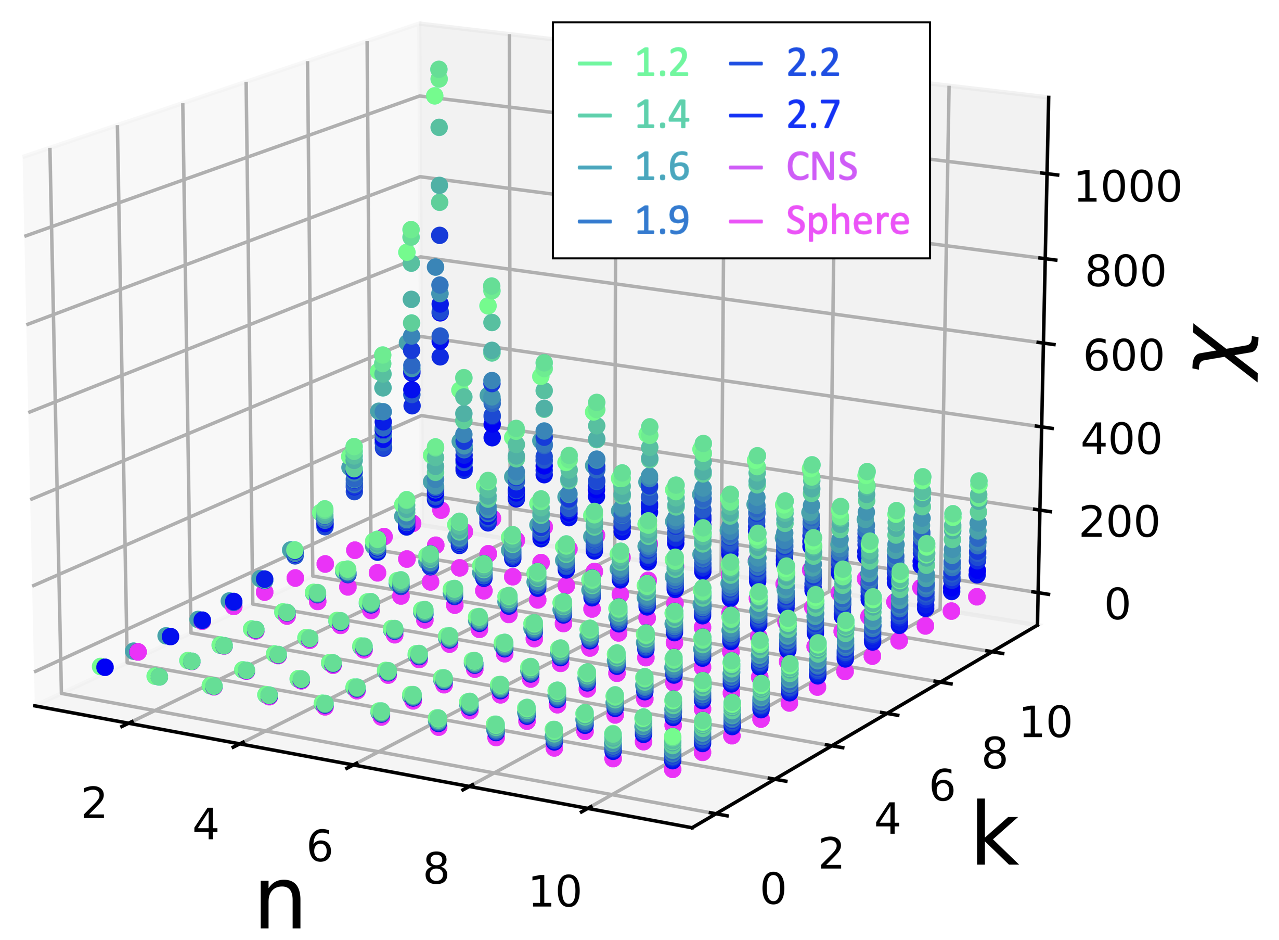}
    \caption{Values of $\chi$ obtained for all fractal aggregates from Fig.~\ref{fig:training_fractals}, at $R=0.1~\mu$m and $\lambda=1,000~\mu$m (fractal dimension $d_f$ shown on the legend). The compact nonspherical (CNS) aggregates and a pseudosphere are shown in pink.}
    \label{fig:all_n_k_data}
\end{figure}

\begin{figure}
    \includegraphics[width=\columnwidth]{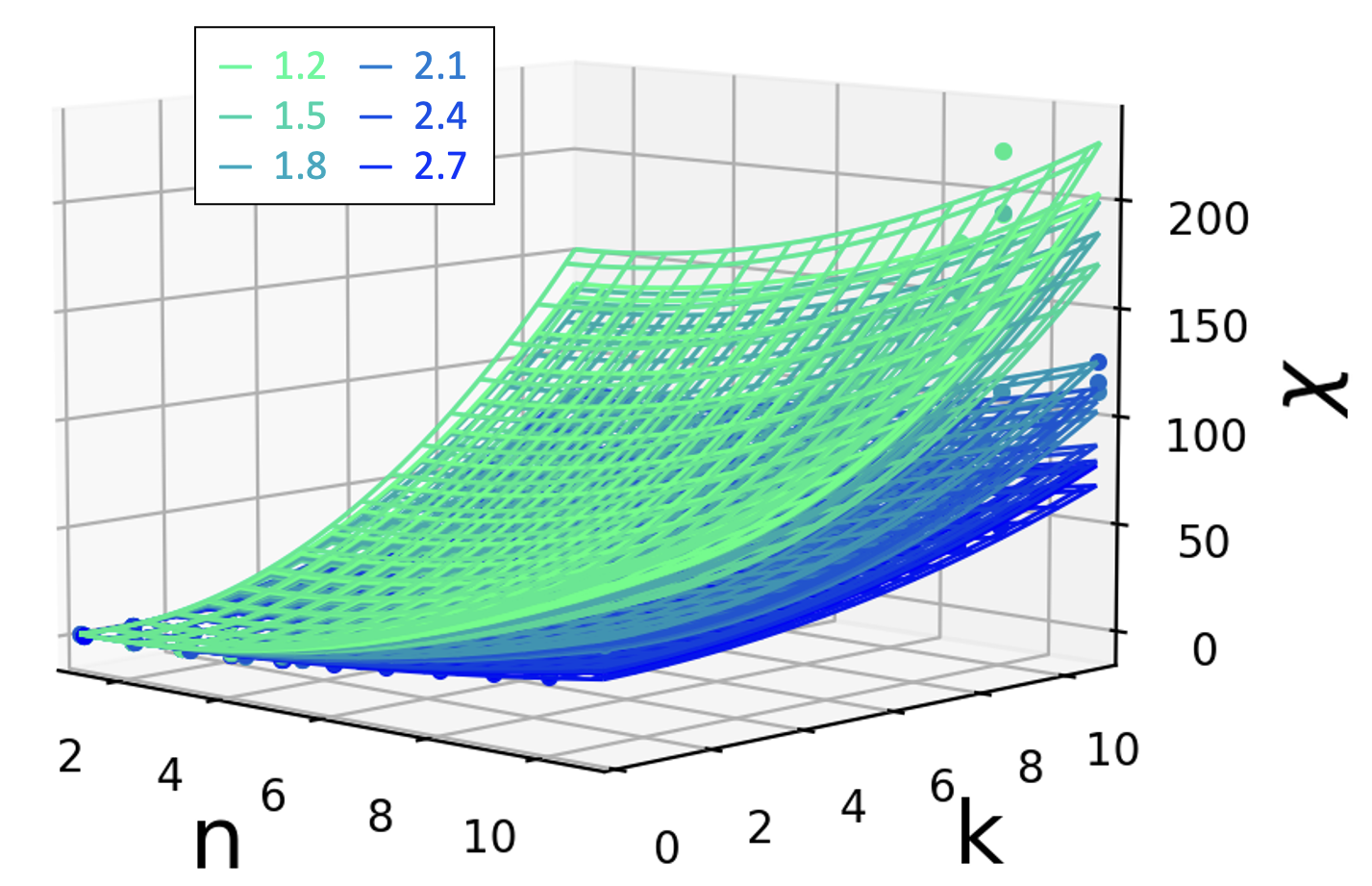}
    \caption{Weighted fit of the multivariate quadratic given by Eq.~\ref{eq:chi_n_k_df} for each fractal aggregate from Fig.~\ref{fig:training_fractals} (fractal dimension $d_f$ shown on the legend). This plot shows the fit for data in the $(n+2) \geq k$ region.}
    \label{fig:quadratic_plot_n_more_than_k}
\end{figure}

\begin{figure}
    \includegraphics[width=\columnwidth]{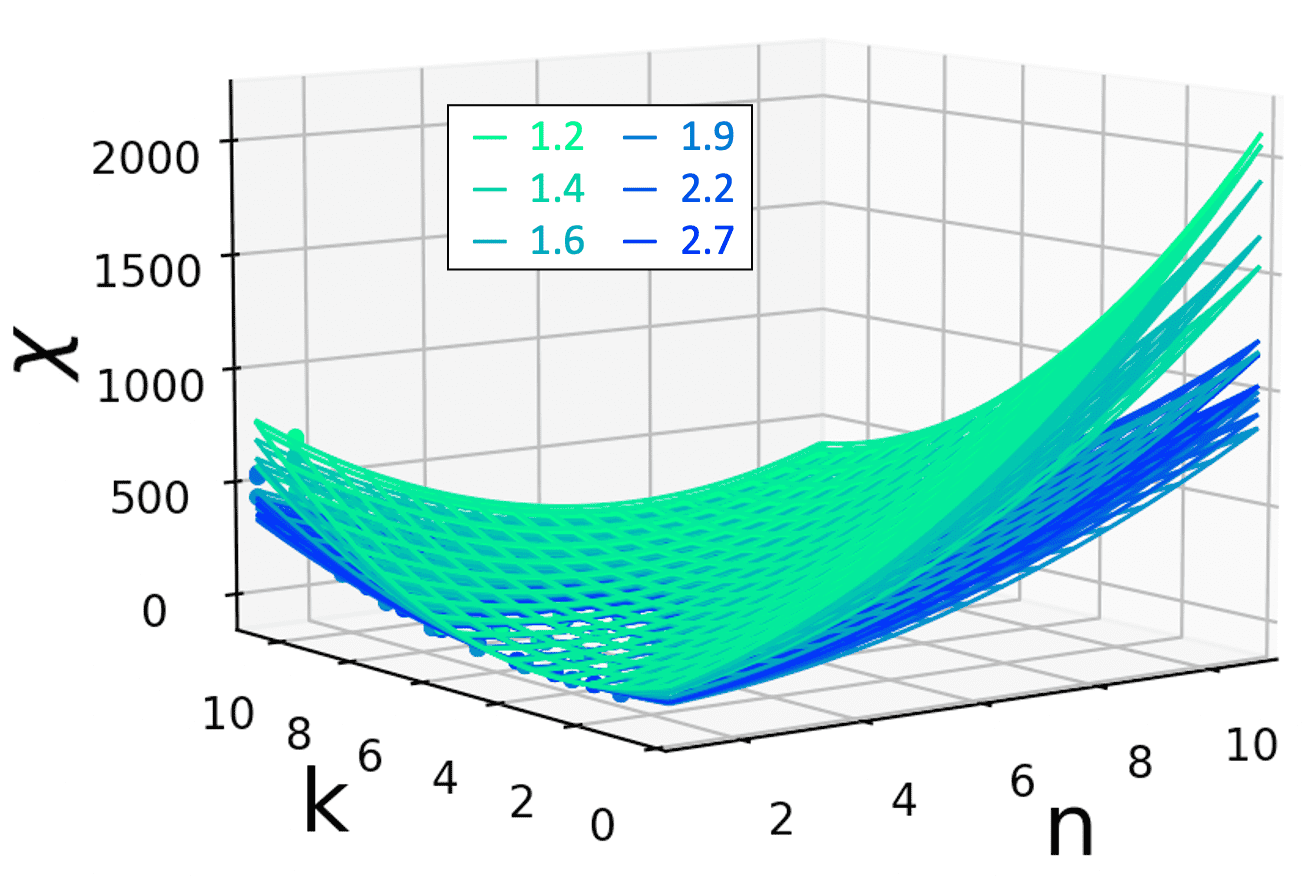}
    \caption{Weighted fit of the multivariate quadratic given by Eq.~\ref{eq:chi_n_k_df} for each fractal aggregate from Fig.~\ref{fig:training_fractals} (fractal dimension $d_f$ shown on the legend). This plot shows the fit for data in the $n<k$ region.}
    \label{fig:quadratic_plot_n_less_than_k}
\end{figure}

\subsection{Fitting a model to the data}

Fig.~\ref{fig:all_n_k_data} shows the enhancement factor $\chi$ plotted as a function of the real and imaginary components of refractive index $(n,k)$. It demonstrates that there is a clear relationship between shape, refractive index and the enhancement factor $\chi(n,k,d_f)$. Long chains of monomers (e.g. $d_f=1.2$) exhibit extreme deviation from the spherical model (enhancement of up to almost three orders of magnitude higher absorption than that predicted by spheres). As the shape becomes more compact (as fractal dimension increases), the aggregates show smaller enhancements in absorption, and the particles that are almost spherical (the CNS and pseudosphere shapes) show virtually no enhancement. After exploring a variety of functional forms that could fit the distribution of $\chi(n,k,d_f)$, we have found that a multivariate quadratic is a good approximation (see Fig.~\ref{fig:quadratic_plot_n_more_than_k} and \ref{fig:quadratic_plot_n_less_than_k}):
\begin{equation} \label{eq:chi_n_k_df}
    \chi(n,k,d_f) = a_{0} + a_{1}n + a_{2}k + a_{3}n^{2} + a_{4}nk + a_{5}k^{2}.
\end{equation}
All of the leading coefficients $a_{0}$ to $a_{5}$ are functions of $d_\mathrm{f}$ (see section \ref{sec:coefficients_a}). At this stage, we highlight that the shapes that were almost completely spherical (CNS and pseudospheres) deviate very little from the spherical model ($\chi<2$ for all refractive indices, shown in Fig.~\ref{fig:all_n_k_data}, as expected). Therefore the original equation for spheres (Eq.~\ref{eq:Rayleigh_original}) gives a good estimate of the absorption efficiency for these shapes at any refractive index, and we omit them from further analysis.

Fitting Eq.~\ref{eq:chi_n_k_df} to the entire $(n,k)$ range of fractal aggregates gives an average $\mathrm{R}^{2}_\mathrm{c}$ (coefficient of determination) value of 0.71. However, if we make the function bimodal, fitting to regions $(n+2) \geq k$ and $(n+2)<k$ separately, we can improve the quality of the fit to 0.99 and 0.92 respectively. This is a significant improvement, and the only additional computation required is a single \lq{if}\rq~statement to determine the region of $(n,k)$ space. We explored other points to divide the two regimes (e.g. $(n+1) \geq k$), but we found the best overall model performance by making the division at $(n+2) \geq k$. We also explored using higher-order functions instead, for example cubic and quartic models, but any small improvement in $\mathrm{R}^{2}_\mathrm{c}$ came at the cost of overfitting the data, which resulted in obscuring the relationship with aggregate shape type (preventing the model from being universally applicable to any shape; see analysis in section~\ref{sec:coefficients_a}). Our philosophy has therefore been to keep the model as simple as possible, whilst ensuring reasonable accuracy ($\approx10\%$).

\subsection{Weighting the fit}

If we fit Eq.~\ref{eq:chi_n_k_df} as least-squares regression naively, the value of $\chi$ can go below 0 for low $(n,k)$ values, which would be an unphysical result. To solve this, we apply a weighting to the low $(n,k)$ values -- a rational choice because the errors in DDA are lowest for values near $m=1+0\mathrm{i}$, and so we have more confidence that these results are the most correct. We experimented with several weighting functions, but found that the following function for assumed error ($\sigma$) ensured a smooth and physical result for low $(n,k)$ values, without sacrificing average accuracy beyond 10\% at higher $(n,k)$:
\begin{equation} \label{eq:sigma}
    \sigma = (n-1)^2 + k^2.
\end{equation}
The form of this function comes from a simple assumption that errors increase as the square of the hypotenuse between value $(n,k)=(1,0)$ and the rest of the grid in $(n,k)$ space. As a result, the average $\mathrm{R}^{2}_\mathrm{c}$ for the $(n+2) \geq k$ region becomes 0.96, and $\mathrm{R}^{2}_\mathrm{c}$ for $(n+2)<k$ becomes 0.91; a small sacrifice in the quality of the fit, but a necessary and worthwhile modelling decision to ensure accurate and physical values for low $(n,k)$ values.

\subsection{Determining the effect of aggregate shape-type} \label{sec:coefficients_a}

Figs.~\ref{fig:quadratic_plot_n_more_than_k} and \ref{fig:quadratic_plot_n_less_than_k} visually demonstrate the weighted fit of Eq.~\ref{eq:chi_n_k_df} to all of the fractal aggregates in Fig.~\ref{fig:training_fractals}, for the regions $(n+2) \geq k$ and $(n+2)<k$ respectively. If we plot the coefficients of $\chi$ as a function of $d_f$ in each case, the correlation is approximately a straight line (see Fig.~\ref{fig:coefficients_n_larger_k} and \ref{fig:coefficients_n_smaller_k}). This is a surprisingly simple result to obtain from such a complicated numerical analysis. Crucially, it allows us to then use the defined straight lines to approximate the coefficients for a fractal aggregate of any shape/fractal dimension, separately for each of the regimes of $(n+2) \geq k$ and $(n+2)<k$:

\begin{figure*}
    \includegraphics[width=\textwidth]{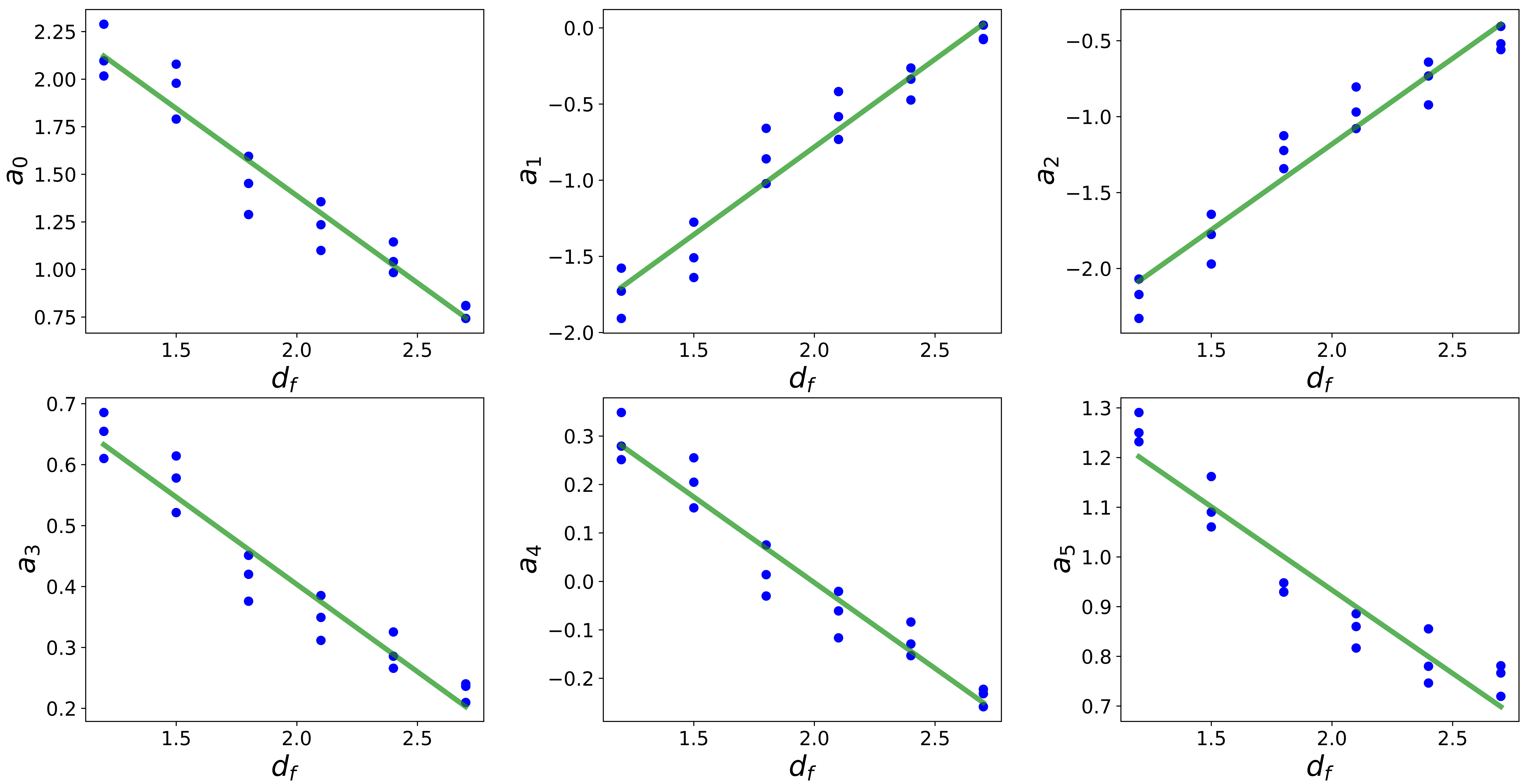}
    \caption{Coefficients found from the weighted fit of Eq.~\ref{eq:chi_n_k_df} to the region each set of data in Fig.~\ref{fig:all_n_k_data} where $(n+2) \geq k$.}
    \label{fig:coefficients_n_larger_k}
\end{figure*}

\begin{figure*}
    \includegraphics[width=\textwidth]{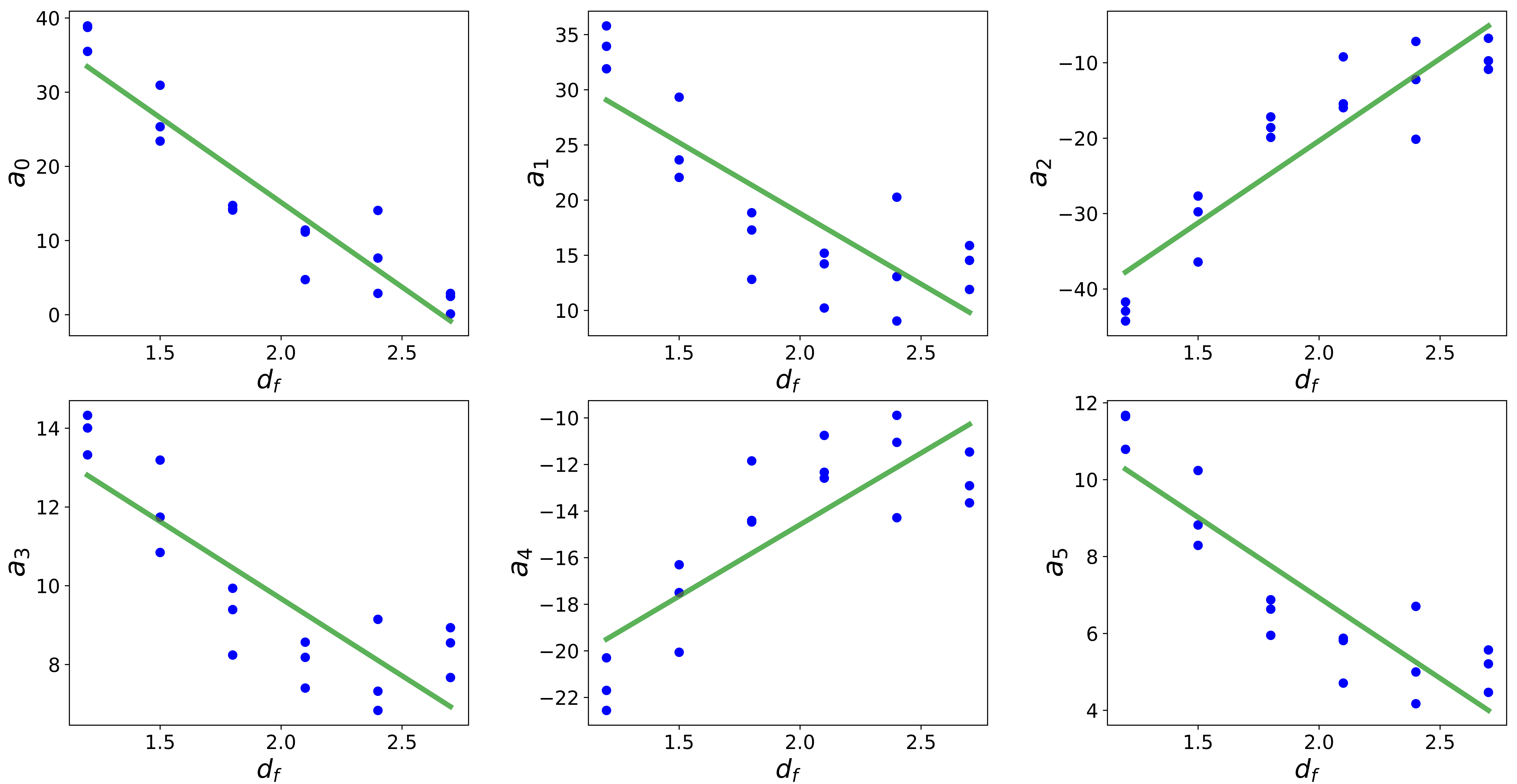}
    \caption{Coefficients found from the weighted fit of Eq.~\ref{eq:chi_n_k_df} to the region each set of data in Fig.~\ref{fig:all_n_k_data} where $(n+2) \leq k$.}
    \label{fig:coefficients_n_smaller_k}
\end{figure*}

\begin{equation} \label{eq:coefficients_n_larger_k}
  \begin{pmatrix}
    a_0 \\
    a_1 \\
    a_2 \\
    a_3 \\
    a_4 \\
    a_5 
  \end{pmatrix}
  =
  \begin{pmatrix}
    -0.917 \\
    1.152 \\
    1.129 \\
    -0.2987 \\
    -0.354 \\
    -0.335 
  \end{pmatrix}
  d_f +
  \begin{pmatrix}
    3.221 \\
    -3.085 \\
    -3.439 \\ 
    0.977 \\
    0.705 \\
    1.604
  \end{pmatrix}
  \mathrm{~for~}(n+2) \geq k
\end{equation}

\begin{equation} \label{eq:coefficients_n_smaller_k}
  \begin{pmatrix}
    a_0 \\
    a_1 \\
    a_2 \\
    a_3 \\
    a_4 \\
    a_5  
  \end{pmatrix}
  =
  \begin{pmatrix}
    -22.844 \\
    -12.818 \\
    21.763 \\
    -3.916 \\
    6.147 \\
    -4.184
  \end{pmatrix}
  d_f +
  \begin{pmatrix}
    60.840 \\ 
    44.436 \\
    -63.879 \\
    17.504 \\
    -26.877 \\ 
    15.293 \\
  \end{pmatrix}
  \mathrm{~for~}(n+2)<k
\end{equation}

As one small additional note, the coefficients above were calculated with a slight overlap into each of the respective opposite regimes to ensure a smooth transition between them (rather than a step-function - see Appendix \ref{appendix:smoothly_combining_the_two_regimes} for more details).

\subsection{Summary of the model: \texttt{MANTA-Ray}}

Using the method above, a user can choose a specific shape type (and thus a value for $d_f$), and then calculate the coefficients using Eqs.~\ref{eq:coefficients_n_larger_k} and \ref{eq:coefficients_n_smaller_k}. These coefficients can then be substituted into the multivariate quadratic Eq.~\ref{eq:chi_n_k_df} to find the enhancement, and combined with Eq.~\ref{eq:Modified_Rayleigh}, the absorption efficiency for any particle size, refractive index and wavelength can be calculated with ease. The model is valid as long as the particle is in the long-wavelength limit ($\lambda>100R$) and providing that the aggregates are homogeneous in chemical composition with refractive index $1+0.01\mathrm{i} \leq m \leq 11+11\mathrm{i}$. Materials that exhibit $n<1$ are not included in this paper because this region of $(n,k)$ space needs special consideration (outside the scope of this paper). However, see Appendix~\ref{appendix:real_refractive_index_less_than_1} for some initial thoughts on extending the model to accommodate materials with these values.

\subsection{Testing the model}

\begin{figure*}
    \includegraphics[width=\textwidth]{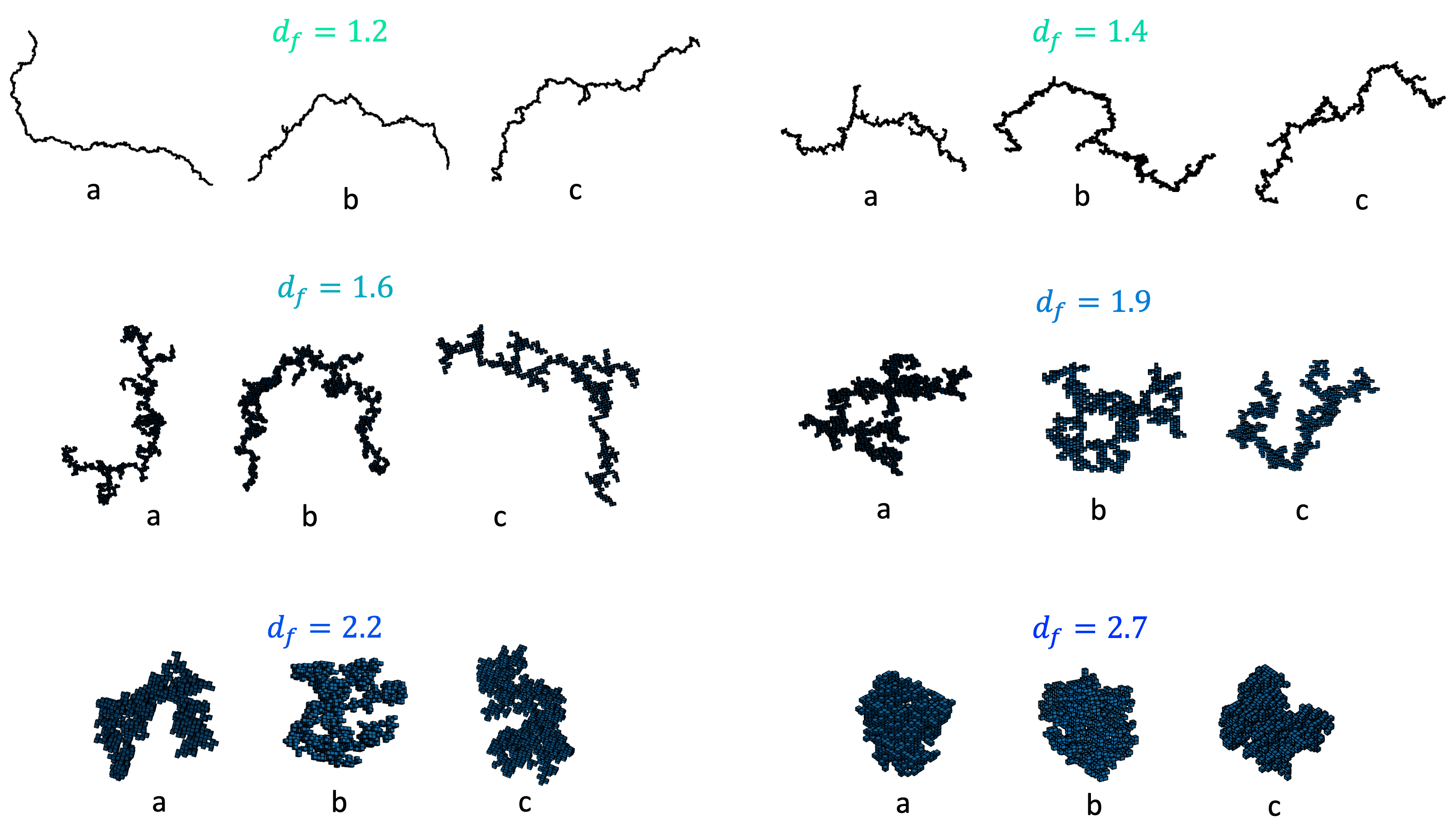}
    \caption{Selection of new aggregates with different fractal dimensions ($d_f$). These aggregates were used to test the model at a different wavelength ($\lambda=100~\mu$m) and radius ($R=0.5~\mu$m) to those that were used to create the model in the first instance ($\lambda=1,000~\mu$m and $R=0.1~\mu$m).}
    \label{fig:test_fractals}
\end{figure*}

To comprehensively test our model, we generated a set of another 18 fractal aggregates of varying fractal dimension between $1.2 \leq d_f \leq 2.7$ (specifically choosing different numerical values to the original dataset -- see Fig.~\ref{fig:test_fractals}). In addition, we altered the wavelength from $1,000~\mu$m to $100~\mu$m, and the particle radius from $0.1~\mu$m to $0.5~\mu$m, to demonstrate the proposition that $\chi(n,k,d_f)$ is independent of particle radius and wavelength (increasing size parameter from $6.28\times10^{-4}$ to $3.14\times10^{-2}$). 

\begin{figure}
    \includegraphics[width=\columnwidth]{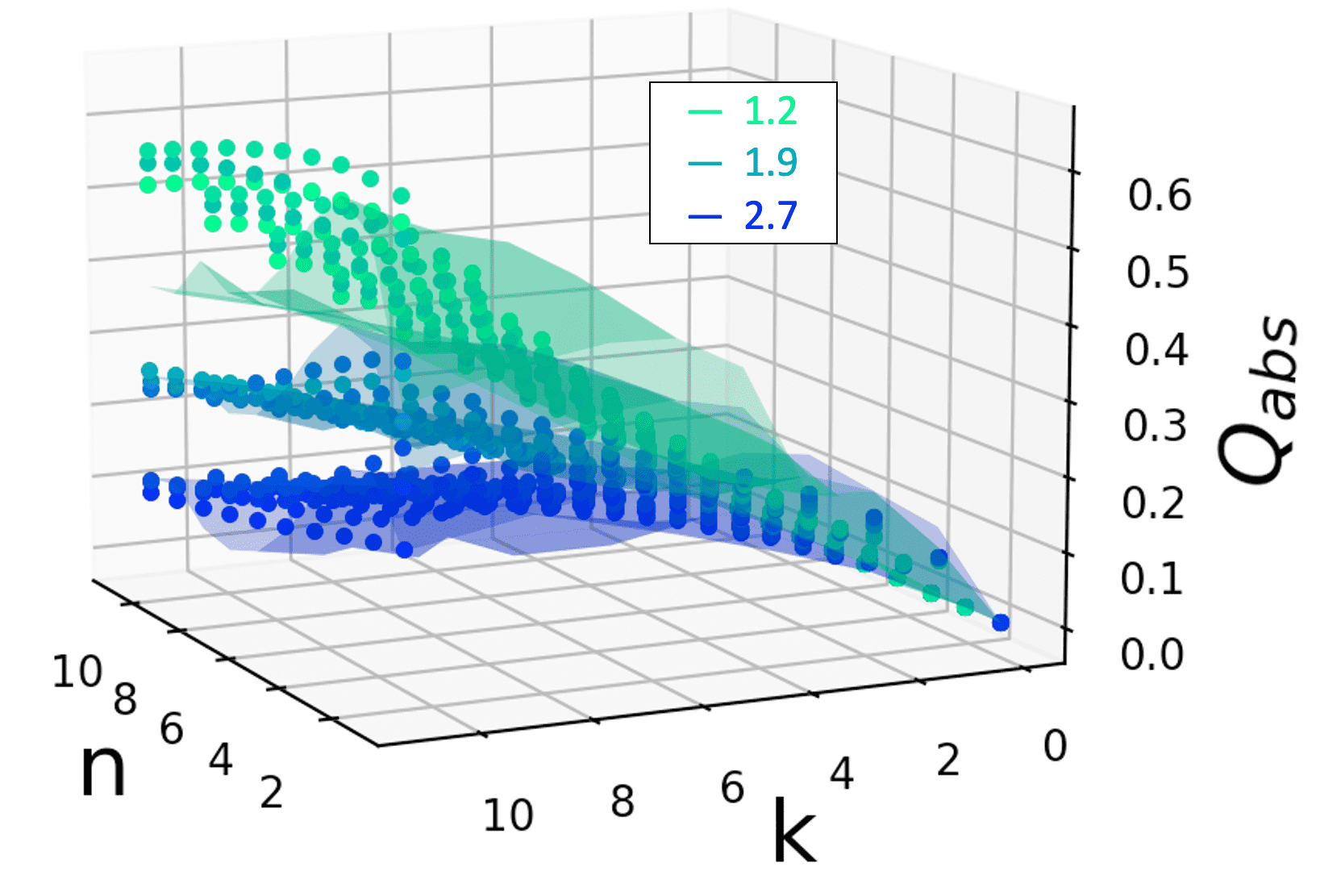}
    \caption{Absorption efficiencies $Q_\mathrm{abs}$ for each of the three shapes (a, b, and c) with fractal dimensions $d_f=$ 1.2, 1.9 and 2.7 (Fig.~\ref{fig:test_fractals}). The results of DDA are shown as circular data points, whilst the \texttt{MANTA-Ray} model fit is shown as the surface (correspondingly coloured). Other fractal dimensions follow an identical pattern, but are omitted here for visual clarity. }
    \label{fig:relative_errors_test_fractals}
\end{figure}

\begin{table*}
    \centering
    \caption{Average and maximum \% errors predicted by modelling particles as spheres ($Q_\mathrm{ext,sphere}$ using Eq.~\ref{eq:Rayleigh_original}) and using \texttt{MANTA-Ray} ($Q_\mathrm{ext,MR}$ using Eq.~\ref{eq:Modified_Rayleigh}) for each test aggregate in Fig.~\ref{fig:test_fractals}. The sample standard deviation is also shown, as a measure of natural variance within aggregates of a particular fractal dimension (shapes defined by $d_f$ still have natural differences in optical properties because of their slight differences in arrangements). This is calculated as an average percentage over all refractive indices in the range $1+0.01\mathrm{i} \leq m \leq 11+11\mathrm{i}$.}
	\label{table:average_and_max_errors}
    \begin{tabular}{c c c r r r r}
    \hline
    \multicolumn{1}{c}{} & \multicolumn{2}{c}{standard deviation within $d_f$} &  \multicolumn{2}{c}{$Q_\mathrm{abs, sphere}$} & \multicolumn{2}{c}{$Q_\mathrm{abs, MR}$} \\
    \cmidrule(lr){2-3} \cmidrule(lr){4-5} \cmidrule(lr){6-7}
    shape   & average (\%) & maximum (\%) & average error (\%) & maximum error (\%) & average error (\%) & maximum error (\%)\\
    \hline
    1.2a &  &  & 12,066.8 & 102,270.5 & 17.9 & 59.8 \\
    1.2b & 3.1 & 6.9 & 12,729.4 & 107,851.1 & 21.1 & 61.1\\
    1.2c &  &  &  12,068.5 & 98,484.9 & 19.9 & 58.7\\
    \hline
    1.4a &  &  & 9,526.4 & 74,558.0 & 15.8 & 51.1 \\
    1.4b & 3.0 & 11.5 & 9,663.4 & 93,669.0 & 14.1 & 51.0\\
    1.4c &  &  &  9,688.2 & 82,207.6 & 14.3 & 51.2\\
    \hline
    1.6a &  &  & 9,320.8 & 77,212.8 & 15.8 & 46.9 \\
    1.6b & 6.1 & 11.8 & 8,431.8 & 72,034.2 & 13.2 & 45.9\\
    1.6c &  &  &  9,654.0 & 90,362.9 & 13.6 & 46.3\\
    \hline
    1.9a &  &  & 7,199.7 & 69,684.3 & 10.9 & 39.5 \\
    1.9b & 3.6 & 10.3 & 6,740.0 & 59,263.2 & 13.4 & 45.1\\
    1.9c &  &  &  7,259.2 & 72,325.2 & 10.7 & 38.7\\
    \hline
    2.2a &  &  & 5,196.1 & 43,351.5 & 15.4 & 60.5 \\
    2.2b & 3.8 & 12.3 & 5,663.1 & 52,457.4 & 12.0 & 42.5\\
    2.2c &  &  &  5,799.3 & 53,703.0 & 10.8 & 47.7\\
    \hline
    2.7a &  &  & 4,758.4 & 53,669.5 & 14.2 & 44.5 \\
    2.7b & 7.3 & 26.1 & 4,389.0 & 45,203.6 & 13.0 & 40.8\\
    2.7c &  &  &  3,779.5 & 31,216.5 & 14.8 & 63.9\\
    \hline
    \end{tabular}
\end{table*}

\begin{figure*}
    \includegraphics[width=\textwidth]{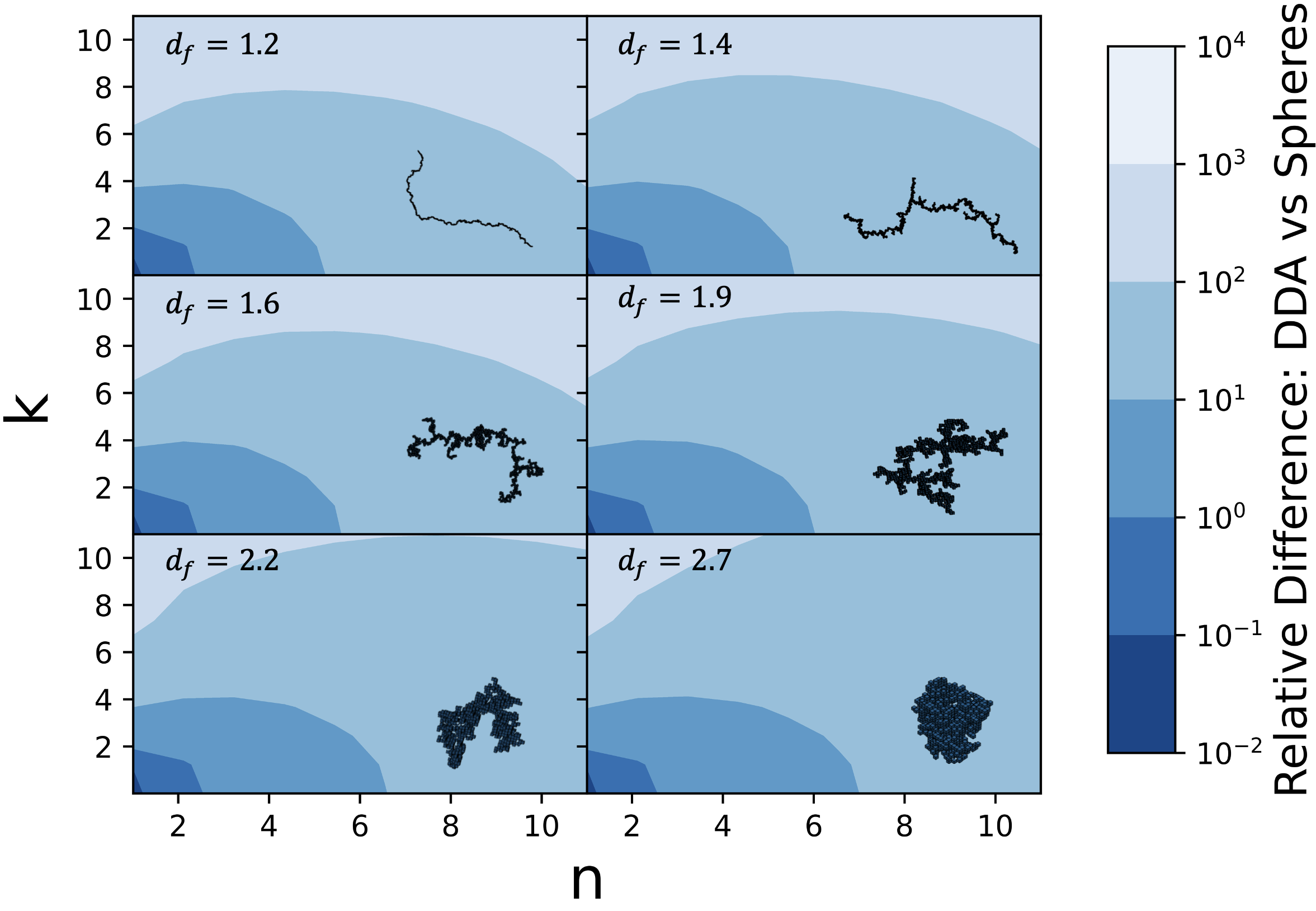}
    \caption{The relative difference in absorption as predicted by DDA versus a sphere of the same mass, for one example of each of the shape types shown above (shape `a' for each fractal dimension from Fig. \ref{fig:test_fractals}). This expands on the values in Table \ref{table:average_and_max_errors} (columns 3 and 4) by showing the relative difference over the entire $(n,k)$ space, as relative differences here (in contrast to the percentages in the table). For example, DDA predicts that a particle with shape $d_\mathrm{f}=1.2$ would exhibit between 1-10 times more absorption than a sphere of the same mass at $(n,k)=(3,2)$.}
    \label{fig:colourmap_spheres}
\end{figure*}

\begin{figure*}
    \includegraphics[width=\textwidth]{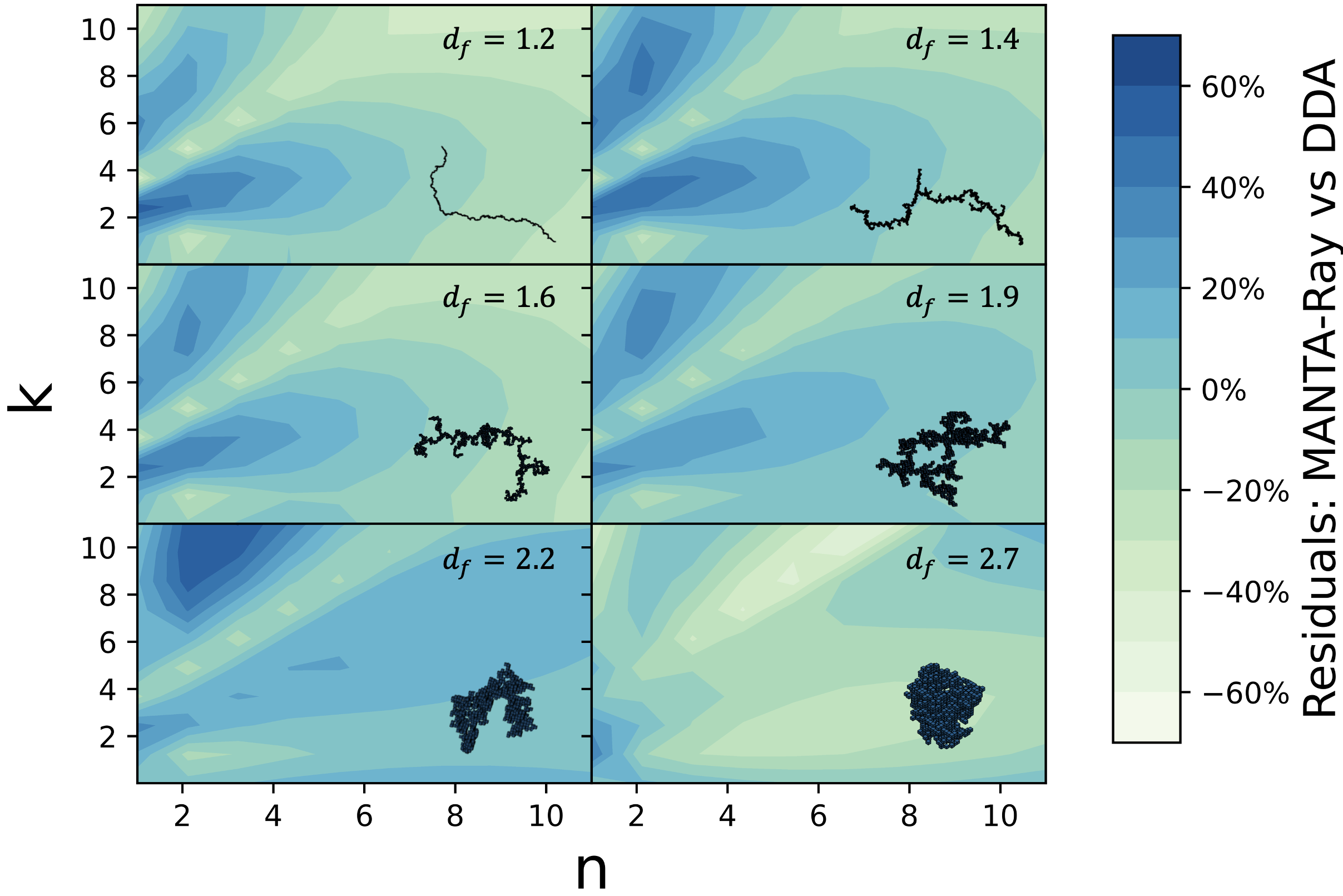}
    \caption{Residuals in the absorption predicted by different particles using \texttt{MANTA-Ray} versus DDA, for one example of each of the shape types shown above (shape `a' for each fractal dimension from Fig. \ref{fig:test_fractals}). This expands on the values in Table \ref{table:average_and_max_errors} (columns 6 and 7) by showing residuals for the entire $(n,k)$ space. The choice of regime change for the bi-modal fit (in this case, $k=n+2$) influences the residuals where the two regions join, and this mode change can be seen visually.}
    \label{fig:colourmap_manta}
\end{figure*}

Table~\ref{table:average_and_max_errors} lists the average and maximum sample standard deviation (including Bessel's correction) for each fractal dimension. This represents the natural variance in optical properties that can be expected because of changes in particle arrangement, even though aggregates may have identical fractal dimension. The compact clusters in particular can have significant variance, which is important to recognise because we are trying to make a predictive model that works for any fractal dimension -- these average and maximum deviation percentages are therefore the absolute minimum errors that any model could be expected to have. In this model we make the assumption that all aggregates with the same fractal dimension have similar structures, however this is not necessarily true. For example, a linear chain made of 4 spheres has a different aspect ratio (and different optical properties) to a linear chain made of 1,000 spheres, despite potentially having an identical fractal dimension \citep[for further discussion of the dependence of absorption on lower $N_\mathrm{mon}$ values, see][]{mackowski2006simplified}. We highlight that this model can be taken to be applicable to shape types that are similar to those shown in Fig.~\ref{fig:training_fractals}, and encourage user-discretion to ensure that this condition is met.

The fourth and fifth columns of Table~\ref{table:average_and_max_errors} list the potential error in absorption one might expect if aggregates are assumed to be spherical (Eq.~\ref{eq:Rayleigh_original}), and the final two columns show the much-reduced error by applying the modification term and using \texttt{MANTA-Ray} (Eq.~\ref{eq:Modified_Rayleigh}). The expected errors from using the spherical or \texttt{MANTA-Ray} models are shown for the full $(n,k)$ parameter space in Figs.~\ref{fig:colourmap_spheres} and \ref{fig:colourmap_manta} respectively. Fig.~\ref{fig:colourmap_spheres} is particularly helpful for estimating the amount that the spherical model might underestimate absorption at specific $(n,k)$ values and for a specific shape type. Errors in \texttt{MANTA-Ray} are typically largest for the largest values of $k$, but these errors are significantly smaller (two orders of magnitude, on average) than if the aggregates are modelled as spheres.

To further demonstrate which combinations of refractive indices and shape types have the largest errors, Fig.~\ref{fig:relative_errors_test_fractals} compares the results obtained by rigorous DDA calculations (data points) and the \texttt{MANTA-Ray} model (surface plot), for shapes of fractal dimension 1.2, 1.9 and 2.7. The largest errors occur for $d_f=1.2$ at the highest refractive indices ($m=11+11\mathrm{i}$). However, even in this region, the errors presented by \texttt{MANTA-Ray} are again much smaller than if the spherical model is used.

\subsection{Assessing Errors}

There are several sources of error that we expect to appear in the final model:

\begin{enumerate}
    \item The fit of the multivariate quadratic Eq.~\ref{eq:chi_n_k_df} to each set of data is good, but not perfect. We find a maximum of 10\% error for all shapes studied here.
    \item The linear fits used to calculate coefficients for a general model (for any $d_f$) in Fig.~\ref{fig:coefficients_n_larger_k} and \ref{fig:coefficients_n_smaller_k} have $\mathrm{R}^{2}_\mathrm{c}$ (coefficient of determination) values of 0.86 and 0.66 respectively (there is larger variance in the coefficients for shapes of the same fractal dimension). This variance contributes to error in the final model, in the $n<k$ region especially.
    \item DDA is a numerical approximation, and although efforts have been to minimise errors by fulfilling the necessary criteria and ensuring convergence, some error is expected in the results (the exact amount is very difficult to quantify). There are shape errors, discretisation errors, and in addition there are errors in averaging over orientations and polarisation states (though tests have indicated that the latter are no more than a few \%). We suggest \citet{yurkin2006convergence} for further discussion of these.
\end{enumerate}

Considering the above potential sources of error, in addition to the natural variance indicated by the standard deviations in Table \ref{table:average_and_max_errors}, the average errors in Table~\ref{table:average_and_max_errors} are actually reasonably good. In real-use cases, we very rarely know the exact fractal dimension or shapes that aerosols are forming, and so \texttt{MANTA-Ray} works well at estimating absorption here.

As one final comment on the accuracy of the model, it is interesting to point out that the second wavelength, particle size and refractive index combination studied for testing the model is actually beyond the usual region of applicability of the original absorption equation for spherical particles (Eq.~\ref{eq:Rayleigh_original}). Interestingly, we have found that \texttt{MANTA-Ray} still predicts correct values at these size parameters for fractal aggregates (we analyse this in more detail in Appendix~\ref{appendix:keeping_additional_expansion_terms}).

\subsection{The packaged model: \texttt{MANTA-Ray}}
The model developed here has been written and optimised as a downloadable function, written in python; it is freely available at \url{https://github.com/mglodge/MANTA-Ray}. It has been optimised for speed, and simply requires four inputs to predict absorption efficiencies -- fractal dimension ($d_f$), wavelength ($\lambda$), refractive index ($m$), and equivalent-volume spherical radius ($R$). 

\section{Conclusions} \label{section:conclusions}

In summary, in this paper we have presented the following:

\begin{enumerate}

    \item We provide a very powerful new analytical equation (Eq.~\ref{eq:Modified_Rayleigh}) that allows extremely fast calculations for non-spherical fractal aggregates in the long-wavelength limit.
    
    \item We find that the enhanced absorption in fractal aggregates (versus spherical particles of the same mass) is inversely proportional to wavelength in the long-wavelength limit, and that the enhancement factor can be characterised as a simple multiple of the equation for calculating absorption within spheres (see Fig.~\ref{fig:DDA_vs_Mie_vs_Rayleigh} and Table~\ref{table:proof_of_concept}).
    
    \item We determine that the enhancement factor is well described by the multivariate quadratic equation $\chi(n,k,d_f$) (Eq.~\ref{eq:chi_n_k_df}), which is a function of the real and imaginary components of refractive index and the fractal dimension of the aggregate. This model (\texttt{MANTA-Ray}) is valid for any particle composition, fractal aggregate shape, and wavelength within the long-wavelength limit.

    \item This proposed model can be expected to obtain absorption cross-sections with average errors of 10-20\% and maximum errors of 40-70\% (compared to DDA) in the regime of $1 + 0.01\mathrm{i} \leq m \leq 11 + 11\mathrm{i}$. This is a substantial improvement to the average and maximum errors expected if the aggregates are modelled as spheres (3,000--12,000\% and 31,000--110,000\% respectively -- see Table \ref{table:average_and_max_errors}).

    \item The model is $10^{13}$ times faster than DDA at obtaining results within this region.

    \item We provide optimised and packaged code\footnote{\url{https://github.com/mglodge/MANTA-Ray}} (\texttt{MANTA-Ray}) to calculate absorption efficiencies using this model for fractal aggregates in this regime.

    \item We suggest that laboratory studies are performed to experimentally verify the enhancement predicted by DDA for materials with high refractive indices.
    
\end{enumerate}

\texttt{MANTA-Ray} offers a fast and simple way to estimate the enhanced absorption in fractal aggregates predicted by DDA, which may lead to significantly different estimates of radiative forcing (and thus atmospheric temperature) in planetary models, as well as different particle masses/quantities when used in protoplanetary disc models, versus cases where the particles are modelled as spheres. \texttt{MANTA-Ray} significantly speeds up the calculation of optical properties, allowing (for example) the exploration of a larger selection of forward models, by reducing the complex analysis of DDA into a simple analytical function. The simplicity of the equation can also guide a more intuitive understanding of how these aggregates interact with radiation much larger than their characteristic lengths. We encourage integration with existing astrophysical and optical models to explore the potential effects of this enhancement in the Rayleigh regime, in addition to further lab studies to verify the substantial increases in absorption suggested by DDA at the largest refractive indices.

\section*{Acknowledgements}

We thank an anonymous referee, whose comments significantly improved this paper. We thank M. Yurkin for an incredibly helpful discussion about the enhanced absorption effect predicted by DDA, and for providing an analytical proof of the enhanced absorption effect in elongated spheroids. The authors thank D. Grant and P. Carter, for stimulating ideas and discussions which helped formulate the ideas presented here. We thank M. Min for insightful comments on increasing the range of refractive indices studies, to include more extreme cases and therefore widen the applicability of the model to other compositions. ML would like to acknowledge the generous support of the Keith Burgess Scholarship and Frederick Frank Fund that allowed this research to be carried out. Additionally, we thank the authors of \texttt{DDSCAT} \citep{draine1994discrete}, \texttt{ADDA} \citep{yurkin2021capabilities}, \texttt{matplotlib} \citep{Hunter:2007}, \texttt{NumPy}\citep{Hunter:2007} and \texttt{SciPy} \citep{2020SciPy-NMeth} for making their software freely available. We thank B. Draine in particular for making the data files for BAM2 shapes publicly available. HRW and ZML acknowledge the financial support from the Science and Technologies Facilities Council grant number ST/V000454/1. HRW was also funded by UK Research and Innovation (UKRI) under the UK government’s Horizon Europe funding guarantee for an ERC STG award [grant number EP/Y006313/1]. This work was carried out using the computational facilities of the Advanced Computing Research Centre, University of Bristol - http://www.bristol.ac.uk/acrc/.
%%%%%%%%%%%%%%%%%%%%%%%%%%%%%%%%%%%%%%%%%%%%%%%%%%
\section*{Data Availability}

 The \texttt{MANTA-Ray} code is freely available for use at: \url{https://github.com/mglodge/MANTA-Ray}. The code used to perform the DDA calculations (CORAL), and the shape files used in this study are provided as supplementary materials here: \url{https://doi.org/10.5281/zenodo.14002327}.



\bibliographystyle{mnras}
\bibliography{references} % if your bibtex file is called example.bib

\begin{thebibliography}{}
\makeatletter
\relax
\def\mn@urlcharsother{\let\do\@makeother \do\$\do\&\do\#\do\^\do\_\do\%\do\~}
\def\mn@doi{\begingroup\mn@urlcharsother \@ifnextchar [ {\mn@doi@} {\mn@doi@[]}}
\def\mn@doi@[#1]#2{\def\@tempa{#1}\ifx\@tempa\@empty \href {http://dx.doi.org/#2} {doi:#2}\else \href {http://dx.doi.org/#2} {#1}\fi \endgroup}
\def\mn@eprint#1#2{\mn@eprint@#1:#2::\@nil}
\def\mn@eprint@arXiv#1{\href {http://arxiv.org/abs/#1} {{\tt arXiv:#1}}}
\def\mn@eprint@dblp#1{\href {http://dblp.uni-trier.de/rec/bibtex/#1.xml} {dblp:#1}}
\def\mn@eprint@#1:#2:#3:#4\@nil{\def\@tempa {#1}\def\@tempb {#2}\def\@tempc {#3}\ifx \@tempc \@empty \let \@tempc \@tempb \let \@tempb \@tempa \fi \ifx \@tempb \@empty \def\@tempb {arXiv}\fi \@ifundefined {mn@eprint@\@tempb}{\@tempb:\@tempc}{\expandafter \expandafter \csname mn@eprint@\@tempb\endcsname \expandafter{\@tempc}}}

\bibitem[\protect\citeauthoryear{Adachi, Chung  \& Buseck}{Adachi et~al.}{2010}]{adachi2010shapes}
Adachi K.,  Chung S.~H.,   Buseck P.~R.,  2010, Journal of Geophysical Research: Atmospheres, 115

\bibitem[\protect\citeauthoryear{Bazell \& Dwek}{Bazell \& Dwek}{1990}]{bazell1990effects}
Bazell D.,  Dwek E.,  1990, Astrophysical Journal, Part 1 (ISSN 0004-637X), vol. 360, Sept. 1, 1990, p. 142-150., 360, 142

\bibitem[\protect\citeauthoryear{Bohren \& Huffman}{Bohren \& Huffman}{2008}]{bohren2008absorption}
Bohren C.~F.,  Huffman D.~R.,  2008, Absorption and scattering of light by small particles.
John Wiley \& Sons

\bibitem[\protect\citeauthoryear{Bohren \& Singham}{Bohren \& Singham}{1991}]{bohren1991backscattering}
Bohren C.~F.,  Singham S.~B.,  1991, Journal of Geophysical Research: Atmospheres, 96, 5269

\bibitem[\protect\citeauthoryear{Draine}{Draine}{2016}]{draine2016graphite}
Draine B.,  2016, The Astrophysical Journal, 831, 109

\bibitem[\protect\citeauthoryear{Draine \& Flatau}{Draine \& Flatau}{1994}]{draine1994discrete}
Draine B.~T.,  Flatau P.~J.,  1994, Josa a, 11, 1491

\bibitem[\protect\citeauthoryear{Farias, K{\"o}yl{\"u}  \& Carvalho}{Farias et~al.}{1996}]{farias1996range}
Farias T.~L.,  K{\"o}yl{\"u} {\"U}.~{\"O}.,   Carvalho M. d.~G.,  1996, Applied optics, 35, 6560

\bibitem[\protect\citeauthoryear{Filippov, Zurita  \& Rosner}{Filippov et~al.}{2000}]{filippov2000fractal}
Filippov A.,  Zurita M.,   Rosner D.~E.,  2000, Journal of colloid and interface science, 229, 261

\bibitem[\protect\citeauthoryear{Fogel \& Leung}{Fogel \& Leung}{1998}]{fogel1998modeling}
Fogel M.~E.,  Leung C.~M.,  1998, The Astrophysical Journal, 501, 175

\bibitem[\protect\citeauthoryear{Gao, Wakeford, Moran  \& Parmentier}{Gao et~al.}{2021}]{gao2021aerosols}
Gao P.,  Wakeford H.~R.,  Moran S.~E.,   Parmentier V.,  2021, Aerosols in exoplanet atmospheres

\bibitem[\protect\citeauthoryear{Gay-Balmaz \& Martin}{Gay-Balmaz \& Martin}{2002}]{gay2002library}
Gay-Balmaz P.,  Martin O.~J.,  2002, Computer physics communications, 144, 111

\bibitem[\protect\citeauthoryear{Henning \& Semenov}{Henning \& Semenov}{2013}]{henning2013chemistry}
Henning T.,  Semenov D.,  2013, Chemical Reviews, 113, 9016

\bibitem[\protect\citeauthoryear{Henning, Michel  \& Stognienko}{Henning et~al.}{1995}]{henning1995dust}
Henning T.,  Michel B.,   Stognienko R.,  1995, Planetary and Space Science, 43, 1333

\bibitem[\protect\citeauthoryear{Hinsen \& Felderhof}{Hinsen \& Felderhof}{1992}]{hinsen1992dielectric}
Hinsen K.,  Felderhof B.,  1992, Physical Review B, 46, 12955

\bibitem[\protect\citeauthoryear{Hoshyaripour, Bachmann, F{\"o}rstner, Steiner, Vogel, Wagner, Walter  \& Vogel}{Hoshyaripour et~al.}{2019}]{hoshyaripour2019effects}
Hoshyaripour G.,  Bachmann V.,  F{\"o}rstner J.,  Steiner A.,  Vogel H.,  Wagner F.,  Walter C.,   Vogel B.,  2019, Journal of Geophysical Research: Atmospheres, 124, 7164

\bibitem[\protect\citeauthoryear{Hunter}{Hunter}{2007}]{Hunter:2007}
Hunter J.~D.,  2007, \mn@doi [Computing in Science \& Engineering] {10.1109/MCSE.2007.55}, 9, 90

\bibitem[\protect\citeauthoryear{Kahnert, Nousiainen  \& Lindqvist}{Kahnert et~al.}{2014}]{kahnert2014model}
Kahnert M.,  Nousiainen T.,   Lindqvist H.,  2014, Journal of Quantitative Spectroscopy and Radiative Transfer, 146, 41

\bibitem[\protect\citeauthoryear{Katrinak, Rez, Perkes  \& Buseck}{Katrinak et~al.}{1993}]{katrinak1993fractal}
Katrinak K.~A.,  Rez P.,  Perkes P.~R.,   Buseck P.~R.,  1993, Environmental science \& technology, 27, 539

\bibitem[\protect\citeauthoryear{K{\"o}hler, Guillet  \& Jones}{K{\"o}hler et~al.}{2011}]{kohler2011aggregate}
K{\"o}hler M.,  Guillet V.,   Jones A.,  2011, Astronomy \& Astrophysics, 528, A96

\bibitem[\protect\citeauthoryear{Liu, Wong, Snelling  \& Smallwood}{Liu et~al.}{2013}]{liu2013investigation}
Liu F.,  Wong C.,  Snelling D.~R.,   Smallwood G.~J.,  2013, Aerosol Science and Technology, 47, 1393

\bibitem[\protect\citeauthoryear{Liu, Teng, Zhu, Yurkin  \& Yung}{Liu et~al.}{2018}]{liu2018performance}
Liu C.,  Teng S.,  Zhu Y.,  Yurkin M.~A.,   Yung Y.~L.,  2018, Journal of Quantitative Spectroscopy and Radiative Transfer, 221, 98

\bibitem[\protect\citeauthoryear{Lodge, Wakeford  \& Leinhardt}{Lodge et~al.}{2024}]{lodge2024aerosols}
Lodge M.~G.,  Wakeford H.~R.,   Leinhardt Z.~M.,  2024, Monthly Notices of the Royal Astronomical Society, 527, 11113

\bibitem[\protect\citeauthoryear{Mackowski}{Mackowski}{1995}]{mackowski1995electrostatics}
Mackowski D.~W.,  1995, Applied optics, 34, 3535

\bibitem[\protect\citeauthoryear{Mackowski}{Mackowski}{2006}]{mackowski2006simplified}
Mackowski D.~W.,  2006, Journal of Quantitative Spectroscopy and Radiative Transfer, 100, 237

\bibitem[\protect\citeauthoryear{Mackowski \& Mishchenko}{Mackowski \& Mishchenko}{1996}]{mackowski1996calculation}
Mackowski D.~W.,  Mishchenko M.~I.,  1996, JOSA A, 13, 2266

\bibitem[\protect\citeauthoryear{Mahrt, Marcolli, David, Gr{\"o}nquist, Barthazy~Meier, Lohmann  \& Kanji}{Mahrt et~al.}{2018}]{mahrt2018ice}
Mahrt F.,  Marcolli C.,  David R.~O.,  Gr{\"o}nquist P.,  Barthazy~Meier E.~J.,  Lohmann U.,   Kanji Z.~A.,  2018, Atmospheric Chemistry and Physics, 18, 13363

\bibitem[\protect\citeauthoryear{Maxwell-Garnett}{Maxwell-Garnett}{1904}]{maxwell1904xii}
Maxwell-Garnett J.~C.,  1904, Philosophical Transactions of the Royal Society of London. Series A, Containing Papers of a Mathematical or Physical Character, 203, 385

\bibitem[\protect\citeauthoryear{Mie}{Mie}{1908}]{mie1908beitrage}
Mie G.,  1908, Annalen der physik, 330, 377

\bibitem[\protect\citeauthoryear{Min, Hovenier  \& de Koter}{Min et~al.}{2005}]{min2005modeling}
Min M.,  Hovenier J.,   de Koter A.,  2005, Astronomy \& Astrophysics, 432, 909

\bibitem[\protect\citeauthoryear{Min, Hovenier, Dominik, de Koter  \& Yurkin}{Min et~al.}{2006}]{min2006absorption}
Min M.,  Hovenier J.,  Dominik C.,  de Koter A.,   Yurkin M.,  2006, Journal of Quantitative Spectroscopy and Radiative Transfer, 97, 161

\bibitem[\protect\citeauthoryear{Mishchenko}{Mishchenko}{2009}]{mishchenko2009electromagnetic}
Mishchenko M.~I.,  2009, Journal of Quantitative Spectroscopy and Radiative Transfer, 110, 808

\bibitem[\protect\citeauthoryear{Mishchenko, Travis  \& Mackowski}{Mishchenko et~al.}{1996}]{mishchenko1996t}
Mishchenko M.~I.,  Travis L.~D.,   Mackowski D.~W.,  1996, Journal of Quantitative Spectroscopy and Radiative Transfer, 55, 535

\bibitem[\protect\citeauthoryear{Moteki}{Moteki}{2019}]{aggregate_gen}
Moteki N.,  2019, Github, nmoteki/aggregate generator (v1.1)

\bibitem[\protect\citeauthoryear{Mu{\~n}oz, Volten, Hovenier, Veihelmann, Van Der~Zande, Waters  \& Rose}{Mu{\~n}oz et~al.}{2004}]{munoz2004scattering}
Mu{\~n}oz O.,  Volten H.,  Hovenier J.,  Veihelmann B.,  Van Der~Zande W.,  Waters L.,   Rose W.~I.,  2004, Journal of Geophysical Research: Atmospheres, 109

\bibitem[\protect\citeauthoryear{Nousiainen}{Nousiainen}{2009}]{nousiainen2009optical}
Nousiainen T.,  2009, Journal of Quantitative Spectroscopy and Radiative Transfer, 110, 1261

\bibitem[\protect\citeauthoryear{Piller \& Martin}{Piller \& Martin}{1998}]{piller1998increasing}
Piller N.~B.,  Martin O.~J.,  1998, IEEE Transactions on Antennas and Propagation, 46, 1126

\bibitem[\protect\citeauthoryear{Purcell \& Pennypacker}{Purcell \& Pennypacker}{1973}]{purcell1973scattering}
Purcell E.~M.,  Pennypacker C.~R.,  1973, The Astrophysical Journal, 186, 705

\bibitem[\protect\citeauthoryear{Rayleigh}{Rayleigh}{1871}]{rayleigh1871light}
Rayleigh L.,  1871, Phil Mag, 41, 274

\bibitem[\protect\citeauthoryear{Rowe et~al.,}{Rowe et~al.}{2015}]{rowe2015galsim}
Rowe B.~T.,  et~al., 2015, Astronomy and Computing, 10, 121

\bibitem[\protect\citeauthoryear{Shen, Draine  \& Johnson}{Shen et~al.}{2008}]{shen2008modeling}
Shen Y.,  Draine B.,   Johnson E.~T.,  2008, The Astrophysical Journal, 689, 260

\bibitem[\protect\citeauthoryear{Sheu, Kumar  \& Cukier}{Sheu et~al.}{1990}]{sheu1990simulation}
Sheu S.-Y.,  Kumar S.,   Cukier R.,  1990, Physical Review B, 42, 1431

\bibitem[\protect\citeauthoryear{Stognienko, Henning  \& Ossenkopf}{Stognienko et~al.}{1995}]{stognienko1995optical}
Stognienko R.,  Henning T.,   Ossenkopf V.,  1995, Astronomy and Astrophysics, v. 296, p. 797, 296, 797

\bibitem[\protect\citeauthoryear{Tazaki}{Tazaki}{2021}]{Tazaki_2021}
Tazaki R.,  2021, \mn@doi [Monthly Notices of the Royal Astronomical Society] {10.1093/mnras/stab1069}, 504, 2811–2821

\bibitem[\protect\citeauthoryear{Virtanen et~al.,}{Virtanen et~al.}{2020}]{2020SciPy-NMeth}
Virtanen P.,  et~al., 2020, \mn@doi [Nature Methods] {10.1038/s41592-019-0686-2}, \href {https://rdcu.be/b08Wh} {17, 261}

\bibitem[\protect\citeauthoryear{Wakeford \& Sing}{Wakeford \& Sing}{2015}]{wakeford2015transmission}
Wakeford H.~R.,  Sing D.~K.,  2015, Astronomy \& Astrophysics, 573, A122

\bibitem[\protect\citeauthoryear{Waterman}{Waterman}{1965}]{waterman1965matrix}
Waterman P.,  1965, Proceedings of the IEEE, 53, 805

\bibitem[\protect\citeauthoryear{West, Doose, Eibl, Tomasko  \& Mishchenko}{West et~al.}{1997}]{west1997laboratory}
West R.~A.,  Doose L.~R.,  Eibl A.~M.,  Tomasko M.~G.,   Mishchenko M.~I.,  1997, Journal of Geophysical Research: Atmospheres, 102, 16871

\bibitem[\protect\citeauthoryear{Wolf \& Toon}{Wolf \& Toon}{2010}]{wolf2010fractal}
Wolf E.,  Toon O.,  2010, Science, 328, 1266

\bibitem[\protect\citeauthoryear{Wright}{Wright}{1987}]{wright1987long}
Wright E.~L.,  1987, Astrophysical Journal, Part 1 (ISSN 0004-637X), vol. 320, Sept. 15, 1987, p. 818-824., 320, 818

\bibitem[\protect\citeauthoryear{Xu}{Xu}{1995}]{xu1995electromagnetic}
Xu Y.-l.,  1995, Applied optics, 34, 4573

\bibitem[\protect\citeauthoryear{Yon, Lemaire, Therssen, Desgroux, Coppalle  \& Ren}{Yon et~al.}{2011}]{yon2011examination}
Yon J.,  Lemaire R.,  Therssen E.,  Desgroux P.,  Coppalle A.,   Ren K.~F.,  2011, Applied Physics B, 104, 253

\bibitem[\protect\citeauthoryear{Yurkin \& Hoekstra}{Yurkin \& Hoekstra}{2007}]{yurkin2007discrete}
Yurkin M.~A.,  Hoekstra A.~G.,  2007, Journal of Quantitative Spectroscopy and Radiative Transfer, 106, 558

\bibitem[\protect\citeauthoryear{Yurkin, Maltsev  \& Hoekstra}{Yurkin et~al.}{2006}]{yurkin2006convergence}
Yurkin M.~A.,  Maltsev V.~P.,   Hoekstra A.~G.,  2006, JOSA A, 23, 2592

\bibitem[\protect\citeauthoryear{Yurkin, Min  \& Hoekstra}{Yurkin et~al.}{2010}]{yurkin2010application}
Yurkin M.~A.,  Min M.,   Hoekstra A.~G.,  2010, Physical Review E, 82, 036703

\bibitem[\protect\citeauthoryear{Yurkin, Smunev, Glukhova, Kichigin, Moskalensky  \& Inzhevatkin}{Yurkin et~al.}{2021}]{yurkin2021capabilities}
Yurkin M.~A.,  Smunev D.~A.,  Glukhova S.~A.,  Kichigin A.~A.,  Moskalensky A.~E.,   Inzhevatkin K.~G.,  2021, in 2021 Radiation and Scattering of Electromagnetic Waves (RSEMW). pp 111--114

\makeatother
\end{thebibliography}

%\begin{thebibliography}{99}
%\end{thebibliography}

%%%%%%%%%%%%%%%%%%%%%%%%%%%%%%%%%%%%%%%%%%%%%%%%%%

%%%%%%%%%%%%%%%%% APPENDICES %%%%%%%%%%%%%%%%%%%%%

\appendix

\section{Smoothly combining the two regimes} \label{appendix:smoothly_combining_the_two_regimes}

\begin{figure}
    \includegraphics[width=\columnwidth]{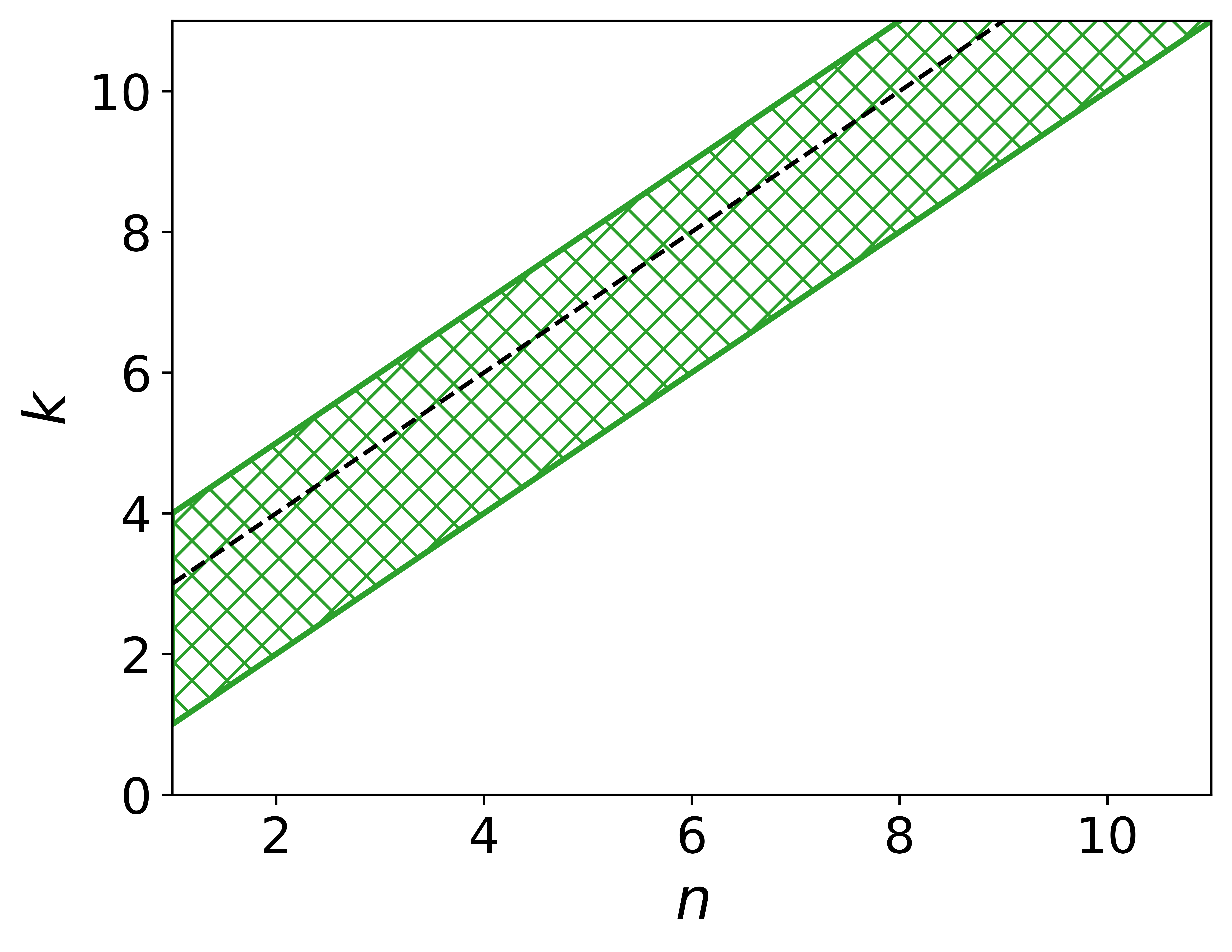}
    \caption{Hatched area shows the region of overlap in $(n,k)$ space that was used to calculate the coefficients for Eq~\ref{eq:chi_n_k_df} for each of the regimes $(n+3) \geq k$ (bottom-right region) and $n \leq k$ (top-left region), to ensure that the bi-modal function has a smooth transition between regimes (suggested transition point shown by the dashed line).}
    \label{fig:n_k_overlap}
\end{figure}

\begin{figure}
    \includegraphics[width=\columnwidth]{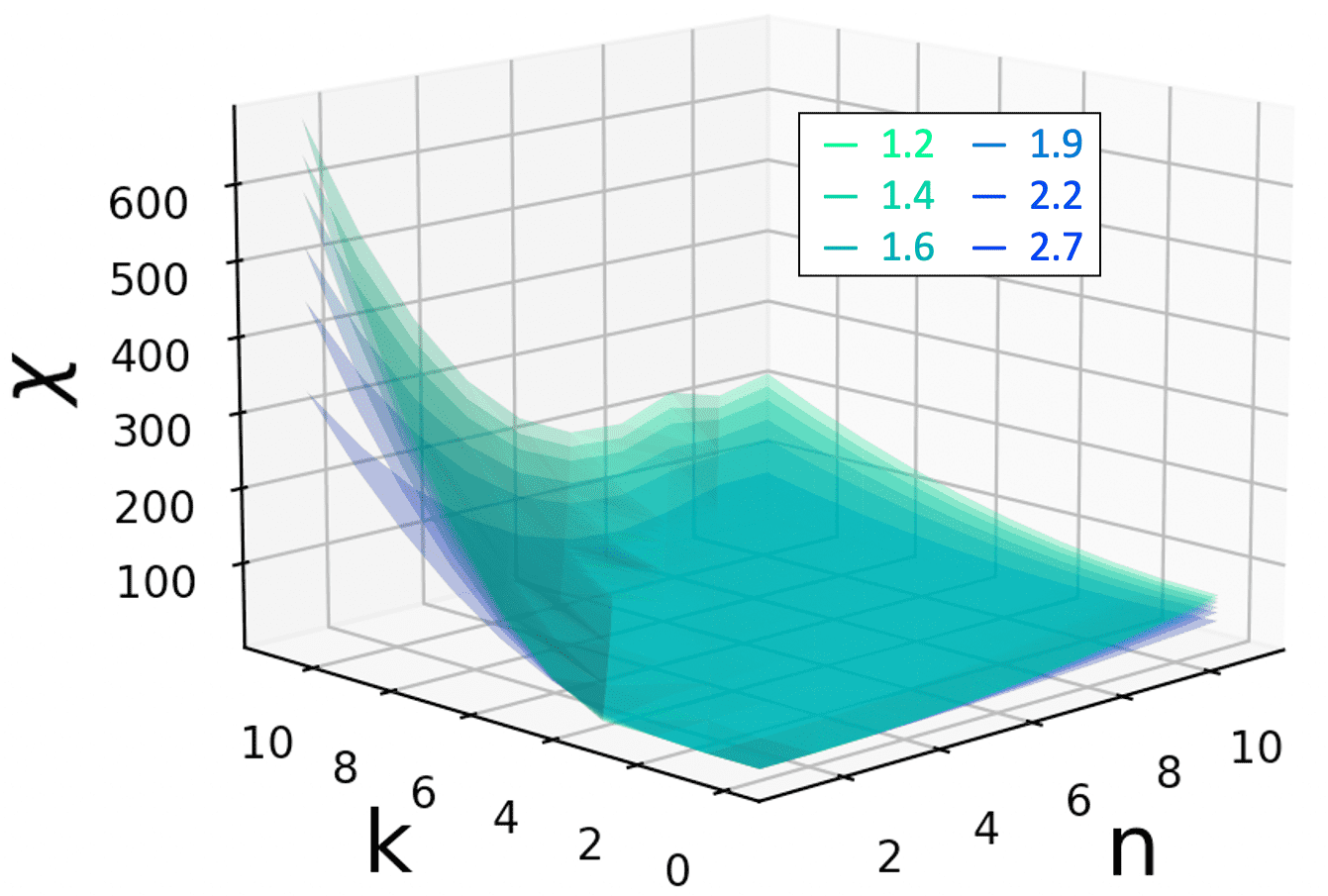}
    \caption{Values of $\chi$ calculated using Eqs.~\ref{eq:coefficients_n_larger_k}, \ref{eq:coefficients_n_smaller_k} and Eq.~\ref{eq:chi_n_k_df} for the test aggregates in Fig.~\ref{fig:test_fractals}. The transition between the regimes $(n+2) \geq k$ and $(n+2) < k$ is smooth as a result of fitting to overlapped $(n,k)$ regions (see Fig.~\ref{fig:n_k_overlap}).}
    \label{fig:final_smooth_model_X}
\end{figure}

To avoid a step-function when transitioning between the two regimes of $(n+2) \geq k$ and $(n+2) < k$, it is important to consider how to connect them. To ensure a smooth final function, we calculated the coefficients in Eq~\ref{eq:chi_n_k_df} for the regions $(n+3) \geq k$ and $n \leq k$ respectively, such that there is some overlap between the two regimes (see Fig.~\ref{fig:n_k_overlap}). This means that the two separate functions for $\chi$ meet more closely along the transition, and once these are used to find the general coefficients in Eq.~\ref{eq:coefficients_n_larger_k} - \ref{eq:coefficients_n_smaller_k}, the result is a much smoother function (see Fig.~\ref{fig:final_smooth_model_X}). Smoothness can be improved further by considering approaches such as "blending" the two functions of $\chi$ around the transition point (e.g. taking a weighted average of each function, with a weighting that smoothly varies depending on which regime is `closest'); these additional modelling decisions are somewhat arbitrary, and so a suggested version has been included in the packaged \texttt{MANTA-Ray} code but omitted from the paper.

\section{Real Refractive Index Less Than 1} \label{appendix:real_refractive_index_less_than_1}

Trial extrapolations of the model have been tested and compared to values obtained from DDA where $n<1$, however care should be taken in this region for a number of reasons. Firstly, whilst low-$n$ refractive indices (e.g. $0.1 + \mathrm{i}$) have been previously tested \citep{yurkin2010application}, refractive indices with extremely low $n$ values and extremely high $k$ values (e.g. $0.1 + 11\mathrm{i}$) have not been as well studied. In tests performed as part of this study, DDA for the pseudosphere shape ($N>70,000$ dipoles) predicted absorption that was 10 times higher than that of a perfect sphere at $m=0.1 + 11\mathrm{i}$, indicating that the number of dipoles needs to be significantly increased for DDA to be accurate in this region (and thus a rigorous analysis for all other shapes is outside of the scope of computing power available for this study). In addition, there is one particular region where resonances occur in spheres, that would require special attention.

\subsection{Resonance region (special conditions)}

\begin{figure*}
    \includegraphics[width=\textwidth]{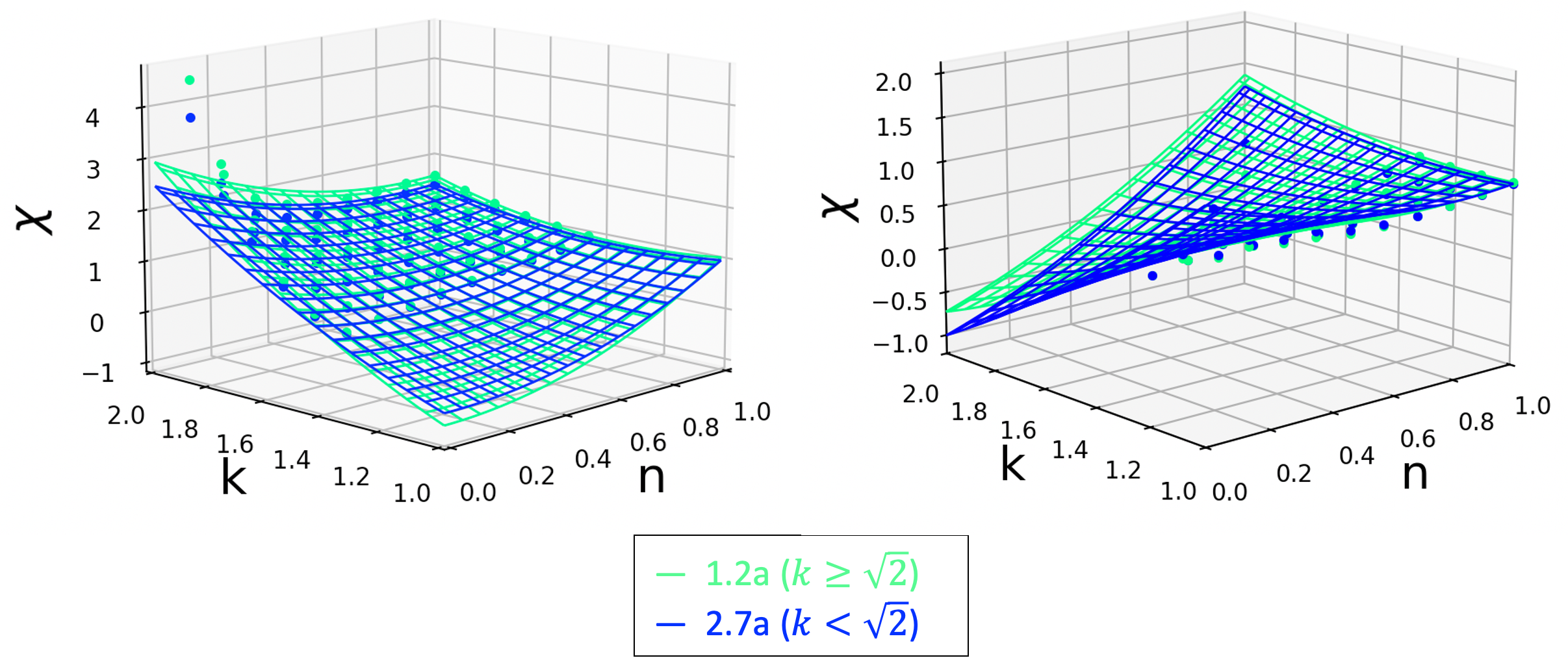}
    \caption{The resonance region, plotted for the two extremes of shape type -- 1.2 (linear) and 2.7 (compact). The quadratic fits and data for $k \geq \sqrt{2}$ and $k<\sqrt{2}$ regions are plotted separately (left and right, respectively) for clarity.}
    \label{fig:resonance_region_1.2_and_2.7}
\end{figure*}

If the model were to be extended for use in the region $n<1$, we highlight that special consideration needs to be made for the resonance region. Although $\chi$ varies smoothly over almost the entire parameter space, for the special case of $m=\sqrt{2}i$ (no real refractive index component), the denominator of Eq.~\ref{eq:Rayleigh_original} becomes 0, creating a strong resonance feature in spherical particles that predicts infinite absorption. This is clearly not a physical result, and so care needs be taken in this region, even when using the original equation and applying it to spheres. However, because our model for non-spherical aggregates (Eq.~\ref{eq:Modified_Rayleigh}) shares the same form, we also need to approach this region with care. For non-spherical particles, sharp resonances are not found because the symmetry of the sphere is broken (as highlighted, for example, by the left panel of Fig.~1 of \citealt{min2005modeling}). To account for this, we suggest to use a function of the same form as $\chi(n,k,d_f)$ in the region between $0+\mathrm{i}\leq m \leq 1+2\mathrm{i}$, but with different coefficients that decrease $\chi$ and \lq{cancel out}\rq~the resonance that would be predicted in this region. At a slight cost of decreasing model elegance by imposing this extra condition, this could expand the regime of applicability of the model to include regions where $n<1$, a property that several materials used in astrophysical models exhibit. Opting to retain as much simplicity as possible, we found that a second bi-modal quadratic function (using Eq.~\ref{eq:chi_n_k_df}, with the modes split at $k=\sqrt{2}$) describes the shape of the resonance region well. As a further simplification, the coefficients were also found to be independent of shape type in this small region. Fig.~\ref{fig:resonance_region_1.2_and_2.7} shows the results for calculations of $\chi$ for the two most different shape types studied in this paper ($d_f=1.2$ and 2.7). The results mostly overlap, and although there are some small differences, once a quadratic has been plotted through the points, the results for the two shapes are show very little variation. In terms of the level of accuracy obtained by the rest of the methodology presented in this paper, we can assume that the functions for this region are independent of shape type.

Assuming that the region is independent of shape type the following formulae for $\chi$ can be used for any fractal aggregate (where $0+i \leq m \leq 1+\sqrt{2}i$):
\begin{equation} \label{eq:resonance_region_1}
    \chi(n,k,d_f) = 4.187 - 3.388n -3.640k + 0.765n^{2} + 2.354nk + 0.591k^{2},
\end{equation}
\noindent and for $0+\sqrt{2}i \leq m \leq 1+2i$:
\begin{equation} \label{eq:resonance_region_2}
    \chi(n,k,d_f) = -1.861 + 3.204n - 0.273k + 1.817n^{2} - 3.186nk + 1.336k^{2}.
\end{equation}
\begin{figure}
    \includegraphics[width=\columnwidth]{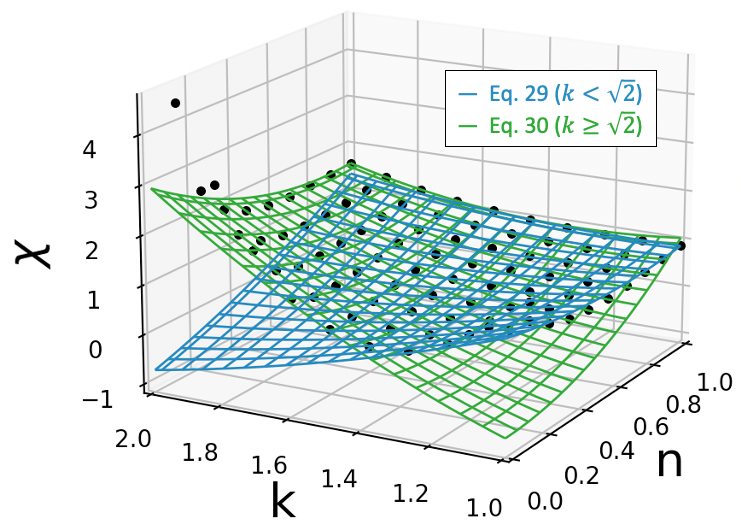}
    \caption{A grid of datapoints with higher resolution (in (n,k) space) near the region where resonance would occur in spheres (but no resonance occurs in non-spherical particles). The fit shown is obtained by using the coefficients from Eq.~\ref{eq:resonance_region_1} (in blue, $k<\sqrt{2}$ region) and \ref{eq:resonance_region_2} (in green, $k \geq \sqrt{2}$ region) substituted into Eq.~\ref{eq:chi_n_k_df}.}
    \label{fig:resonance_region}
\end{figure}

The resulting quadratic plots, including their overlap, are shown in Fig.~\ref{fig:resonance_region}. Care was taken to ensure that both modes intersected with data points at the edge of the region, to avoid a large step when switching between modes. This was achieved by fitting Eqs.~\ref{eq:resonance_region_1} and \ref{eq:resonance_region_2} using a weighted error of $\sigma=\sqrt{1-n}$. For values of $k$ outside of this range ($i<k<2i$, with $n<1$), further studies need to be undertaken.

\section{Keeping additional expansion terms} \label{appendix:keeping_additional_expansion_terms}

If higher-order terms from the Bessel functions are kept in the derivation of Eq.~\ref{eq:Rayleigh_original}, it can be show that \citet{bohren2008absorption}:
\begin{multline} \label{eq:rayleigh_extra_terms} 
    Q_\mathrm{abs,ET} = 4x\mathrm{Im} \left[ \frac{m^2-1}{m^2+2} \left( 1 + \frac{x^2}{15} \left( \frac{m^2-1}{m^2+2} \right) \frac{m^4+27m^2+38}{2m^2+3} \right) \right] + \\ \frac{8}{3}x^4 \mathrm{Re} \left[ \left( \frac{m^2-1}{m^2+2} \right)^2 \right]
\end{multline}
where size parameter $x=\frac{2 \pi R}{\lambda}$, and the \lq{ET}\rq~ denotes \lq{extra terms}\rq. 

For the first set of fractal aggregates that are used to create the model and derive the coefficients in Eq.~\ref{eq:coefficients_n_larger_k}-\ref{eq:coefficients_n_smaller_k} (where $\lambda=1,000~\mu$m and $R=0.1~\mu$m), Eq.~\ref{eq:rayleigh_extra_terms} gives identical results to Eq.~\ref{eq:Rayleigh_original} to 4 significant figures, and the extra terms above are therefore not required.

\begin{figure}
    \includegraphics[width=\columnwidth]{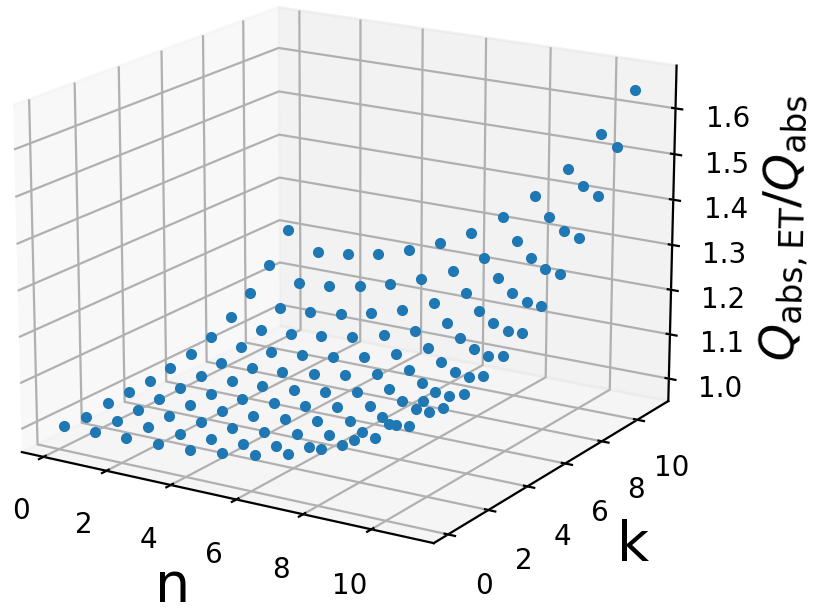}
    \caption{Relative difference between Eq.~\ref{eq:rayleigh_extra_terms} and Eq.~\ref{eq:Rayleigh_original}, expressed as a ratio to highlight the quantitative difference that the extra term can make for high refractive indices at the second size parameter studied in this paper ($\lambda=100~\mu$m and $R=0.5~\mu$m).}
    \label{fig:Rayleigh_original_vs_extra_terms}
\end{figure}

For the second set of fractals studied in this paper (the set that our model is tested on, where $\lambda=100~\mu$m and $R=0.5~\mu$m), at low refractive indices, the two equations are also identical. However, Fig \ref{fig:Rayleigh_original_vs_extra_terms} demonstrates that for the higher refractive indices studied at this higher size parameter, Eq. \ref{eq:Rayleigh_original} can under-predict absorption efficiencies (only $\approx60$\% of the true value). Curiously, our model of applying modification term $\chi(n,k,d_f)$ to the simplified Eq.~\ref{eq:Rayleigh_original} still correctly estimates the true absorption values determined by rigorous DDA analysis for non-spherical shapes, even when the original equation requires more terms to correctly describe the absorption for spheres. We do not provide a rationale for why this is so, but only highlight that it happens, and note that it is useful because it extends the range of the model to larger size parameters (or equivalently, smaller wavelengths).

%%%%%%%%%%%%%%%%%%%%%%%%%%%%%%%%%%%%%%%%%%%%%%%%%%

% Don't change these lines
\bsp	% typesetting comment
\label{lastpage}
\end{document}